\documentclass[aps,amsmath,twocolumn,nofootinbib]{revtex4-2}
\usepackage{graphicx}
\usepackage{relsize}
\usepackage{enumitem}
\usepackage[table,usenames,dvipsnames]{xcolor}
\usepackage{ifthen}
\definecolor{linkcolor}{rgb}{0.6,0,0}
\definecolor{citecolor}{rgb}{0,0,0.75}
\definecolor{urlcolor}{rgb}{0.12,0.46,0.7}
\usepackage[breaklinks, colorlinks, urlcolor=urlcolor, linkcolor=linkcolor,citecolor=citecolor,pdfencoding=auto]{hyperref}
\hypersetup{linktocpage}
\setlist{nolistsep,leftmargin=*} %compress space for all lists

\newcommand{\eg}{e.g.}
\newcommand{\ie}{i.e.}

\def\spose#1{\hbox to 0pt{#1\hss}}
\def\simlt{\mathrel{\spose{\lower 3pt\hbox{$\mathchar"218$}}
     \raise 2.0pt\hbox{$\mathchar"13C$}}}
\def\simgt{\mathrel{\spose{\lower 3pt\hbox{$\mathchar"218$}}
     \raise 2.0pt\hbox{$\mathchar"13E$}}}

\usepackage{xparse,xcoffins}

\ExplSyntaxOn
\NewCoffin\imagecoffin
\NewCoffin\labelcoffin

\keys_define:nn { miguel/label }
 {
  label   .tl_set:N = \l_miguel_label_tl,
  labelbox .bool_set:N = \l_miguel_label_box_bool,
  labelbox .default:n = true,
  fontsize .tl_set:N = \l_miguel_label_size_tl,
  fontsize .initial:n = \footnotesize,
  pos .choice:,
  pos/nw .code:n = \tl_set:Nn \l_miguel_label_pos_tl { left,up },
  pos/ne .code:n = \tl_set:Nn \l_miguel_label_pos_tl { right,up },
  pos/sw .code:n = \tl_set:Nn \l_miguel_label_pos_tl { left,down },
  pos/se .code:n = \tl_set:Nn \l_miguel_label_pos_tl { right,down },
  pos/n .code:n = \tl_set:Nn \l_miguel_label_pos_tl { hc,up },
  pos/w .code:n = \tl_set:Nn \l_miguel_label_pos_tl { left,vc },
  pos/s .code:n = \tl_set:Nn \l_miguel_label_pos_tl { hc,down },
  pos/e .code:n = \tl_set:Nn \l_miguel_label_pos_tl { right,vc },
  pos .initial:n = nw,
  unknown .code:n   = \clist_put_right:Nx \l_miguel_label_clist
                       { \l_keys_key_tl = \exp_not:n { #1 } }
 }
\clist_new:N \l_miguel_label_clist
\box_new:N \l_miguel_label_box
\box_new:N \l_miguel_label_image_box

\NewDocumentCommand{\xincludegraphics}{O{}m}
 {
  \group_begin:
  \tl_clear:N \l_miguel_label_tl
  \clist_clear:N \l_miguel_label_clist
  \keys_set:nn { miguel/label } { #1 }
  \tl_if_empty:NTF \l_miguel_label_tl
   {
    \miguel_includegraphics:Vn \l_miguel_label_clist { #2 }
   }
   {
    \SetHorizontalCoffin\imagecoffin
     {
      \miguel_includegraphics:Vn \l_miguel_label_clist { #2 }
     }
    \SetHorizontalCoffin\labelcoffin
     {
      \raisebox{\depth}
       {
        \bool_if:NTF \l_miguel_label_box_bool
         { \fcolorbox{white}{white}{\l_miguel_label_size_tl\l_miguel_label_tl} }
         { \l_miguel_label_size_tl\l_miguel_label_tl }
       }
     }
    \SetVerticalPole\imagecoffin{left}{3pt+\CoffinWidth\labelcoffin/2}
    \SetVerticalPole\imagecoffin{right}{\Width-3pt-\CoffinWidth\labelcoffin/2}
    \SetHorizontalPole\imagecoffin{up}{\Height-3pt-\CoffinHeight\labelcoffin/2}
    \SetHorizontalPole\imagecoffin{down}{3pt+\CoffinHeight\labelcoffin/2}
    \use:x{\JoinCoffins\imagecoffin[\l_miguel_label_pos_tl]\labelcoffin[vc,hc]} 
    \TypesetCoffin\imagecoffin
   }
   \group_end:
 }
\NewDocumentCommand{\setlabel}{m}
 {
  \keys_set:nn { miguel/label } { #1 }
 }

\cs_new_protected:Nn \miguel_includegraphics:nn
 {
  \includegraphics[#1]{#2}
 }
\cs_generate_variant:Nn \miguel_includegraphics:nn { V }

\ExplSyntaxOff

\begin{document}

\title{Cosmic Microwave Background}
\author{Douglas Scott} \email{dscott@phas.ubc.ca}
\affiliation{University of British Columbia}
\author{George F. Smoot}
\affiliation{HKUST; UC Berkeley; LBNL; DIPC; Paris U.}
\date{August 2023}

\begin{abstract}
This is a review of the current status of studies of the cosmic microwave
background, extracted from Chapter~29 of the 2023 edition of the `Review of
Particle Physics': R.L. Workman et al.\ (Particle Data Group),
Prog.\ Theor.\ Exp.\ Phys., 2022, 083C01 (2022) and 2023 update.
\end{abstract}

\maketitle

\tableofcontents

\section{Introduction}
\label{microwave:sec:CMBintro}

The energy content in electromagnetic radiation from beyond our
Galaxy is dominated by
\index{Background!cosmic microwave, CMB}%
the cosmic microwave background (CMB), discovered in
1965 \cite{microwave:Penzias65,*microwave:Dicke}.
Its spectral distribution is well described by a blackbody
function with $T=2.7255\,$K, which is a principal pillar of the
hot Big Bang model for the Universe, with the
lack of any observed deviations from a Planckian spectrum constraining
physical processes over cosmic history at redshifts $z\lesssim 10^7$
(see earlier versions of this review).

\index{Cosmic background radiation temperature}%
The key information in the CMB sky is extracted from the observed angular
variation of its temperature
(or intensity) correlations, and to a growing extent
polarization \cite{microwave:WSS,*microwave:HD,*microwave:CP}.
After the first detection of CMB anisotropies by the
Cosmic Background
\index{CMB!COBE satellite}%
\index{Satellite!COBE, monopole}%
\index{COBE -- COsmic Background Explorer}%
Explorer ({\it COBE\/}) satellite in 1992 \cite{microwave:smoot92},
there has been intense activity to 
\index{CMB!sky map}%
map the sky at increasing levels of
sensitivity and angular resolution by ground-based and balloon-borne
measurements.  
\index{CMB!Wilkinson Microwave Anisotropy Probe, {\it WMAP}}%
These were joined in 2003 by
the first results from NASA's Wilkinson Microwave Anisotropy Probe
({\it WMAP\/}) \cite{microwave:bennett03},
which were improved upon by analyses of data added every 2 years, culminating
in the 9-year results \cite{microwave:hinshaw12}.
In 2013 we had the first results \cite{microwave:PlanckParams} from the
third generation CMB satellite, ESA's 
\index{Satellite!{\it Planck}, CMB}%
{\it Planck\/} mission \cite{microwave:tauber,*microwave:Planck2013I},
which were enhanced by results from the 2015 {\it Planck\/}
data release \cite{microwave:Planck2015I,microwave:Planck2015Params},
and then the final 2018 {\it Planck\/} data
release \cite{microwave:Planck2018I,microwave:Planck2018Params}.
Additionally, CMB anisotropies have been extended to smaller
angular scales by ground-based experiments, particularly the
Atacama Cosmology Telescope (ACT) \cite{microwave:Swetz11}
and the South Pole Telescope (SPT) \cite{microwave:Carlstrom11}.
Together these observations have led to a stunning confirmation of
the `Standard Model of Cosmology.'  In combination with other astrophysical
data, the CMB anisotropy measurements place quite precise constraints on
a number of cosmological parameters, and have launched us into an era
of precision cosmology.  With more than half a century of study of the CMB,
the program to map temperature anisotropies is effectively wrapping up,
and attention is increasingly focussing on polarization measurements, which
promise further tests of fundamental physics.

\section{CMB Spectrum}
\label{microwave:sec:CMBspectrum}

\index{CMB!spectrum}%
It is well-known that the spectrum of the microwave background is very
precisely that of blackbody radiation, whose temperature evolves with
redshift as $T(z)=T_0(1+z)$ in an expanding Universe.  As a direct confirmation
of its cosmological origin, this relationship has been tested by measuring the
strengths of emission and absorption lines in high-redshift
systems (\eg,~Ref.~\cite{microwave:Klimenko2020,*microwave:Riechers2022}).
%Avoid new limits that come from SZ arguments.

Measurements of the spectrum are
consistent with a blackbody distribution over more than three decades in
frequency (there is a claim by ARCADE \cite{microwave:ARCADE} of a possible
unexpected extragalactic emission signal at low frequency, but the
interpretation is debated \cite{microwave:Singal18}).
All viable cosmological models predict a very nearly
Planckian spectrum to within the current observational limits.
Because of this, measurements of deviations from a blackbody spectrum have
received little attention in recent years, with only a few exceptions.
However, that situation will likely change as proposed
experiments \cite{microwave:PIXIE,*microwave:PRISM,*microwave:Delabrouille21}
are built that have the potential to dramatically
improve early Universe energy release constraints.
It now seems feasible to probe spectral distortion mechanisms that are
{\it required\/} in the standard picture, such as those arising from the
damping and dissipation of relatively small-scale primordial perturbations, or
the average effect of inverse Compton scattering.  A more
ambitious goal would be to reach the precision needed to detect the residual
lines from the cosmological recombination of hydrogen and helium and hence
to test whether conditions at $z\gtrsim1000$ accurately follow those in the
standard picture \cite{microwave:Desjacques15}.

\section{Description of CMB Anisotropies}
\index{Anisotropy of CMB}%
\index{CMB!anisotropy power spectrum}%

Observations show that the CMB contains temperature anisotropies at the
$10^{-5}$ level and polarization anisotropies at the $10^{-6}$ (and lower)
level, over a wide range of angular scales.
These anisotropies are usually expressed using a spherical harmonic
expansion of the CMB sky:
\begin{equation}
\label{microwave:eq:alm}
T(\theta,\phi) = \sum_{\ell m} a_{\ell m} Y_{\ell m}(\theta, \phi)
\end{equation}
(with the linear polarization pattern written in a similar way using the
so-called spin-2 spherical harmonics).
Increasing angular resolution
requires that the expansion goes to higher multipoles.
Because only very weak phase correlations are observed in the CMB sky
and no preferred direction is seen,
the vast majority of the cosmological information is found in the
temperature 2-point function, \ie,~the variance as a function only of angular
separation.  Equivalently, the anisotropy power per unit $\ln\ell$ is
$\ell\sum_m\left|a_{\ell m}\right|^2/4\pi$.

\subsection{The Monopole}

\index{CMB!mean temperature, monopole}%
\index{CMB!monopole}%
The CMB has a mean temperature of $T_\gamma = 2.7255\pm0.0006\,$K
($1\sigma$) \cite{microwave:fixsen09},
which can be considered as the monopole component of CMB maps, $a_{00}$.
Because all mapping experiments involve
difference measurements, they are insensitive to this average level;
monopole measurements can only be made with absolute temperature
devices, such as the FIRAS instrument on the 
{\it COBE\/} satellite \cite{microwave:mather99}.  The measured $k T_\gamma$ is
equivalent to $0.234\,$meV or $4.60\times10^{-10}\,m_{\rm e}c^2$.
A blackbody of the measured temperature
has a number density $n_{\gamma} = (2\zeta(3)/\pi^2)\, T_\gamma^3 \simeq 411\,
{\rm cm^{-3}}$, energy density
$\rho_{\gamma} = (\pi^2 /15)\, T_\gamma^4 \simeq 4.64
 \times 10^{-34}\, {\rm g}\,{\rm cm^{-3}}
 \simeq 0.260\,{\rm eV}\,{\rm cm^{-3}}$, and a fraction of the critical
density ${\rm \Omega}_\gamma\simeq5.38\times10^{-5}$.

\subsection{The Dipole}
\index{CMB!dipole}%

The largest anisotropy is in the $\ell=1$ (dipole) first spherical harmonic,
with amplitude $3.3621\pm0.0010\,$mK \cite{microwave:Planck2018I}.
The dipole is interpreted to be the result of the Doppler boosting of the
monopole caused by the 
\index{CMB!Solar System motion}%
Solar System motion relative to the nearly isotropic
blackbody field, as broadly confirmed by measurements of the radial velocities
of local galaxies (\eg,~Ref.~\cite{microwave:Hoffman15}); the intrinsic
(non-Doppler) part of the signal is expected to be 2 orders of magnitude
smaller (and fundamentally difficult to distinguish).
The motion of an observer with velocity
$\beta \equiv v/c$ relative to an isotropic Planckian radiation field of
temperature ${T_0}$ produces a Lorentz-boosted temperature pattern
\index{CMB!Lorentz-boosted temperature pattern}%
%\begin{align}
\begin{align}
\label{microwave:eq:Toftheta}
T(\theta) &= T_0 (1 - \beta^{2})^{1/2}/(1 - \beta \cos\theta) \cr
%T(\theta) = T_0 (1 - \beta^{2})^{1/2}/(1 - \beta \cos\theta) 
 &\simeq T_0 \, \left[1 + \beta \cos\theta + \left(\beta^{2}/2\right)
% \simeq T_0 \, \left[1 + \beta \cos\theta + \left(\beta^{2}/2\right)
 \cos2\theta + {\rm O}\left(\beta^3\right)\right].
\end{align}
%\end{align}
At every point in the sky, one observes a blackbody spectrum, with
temperature $T(\theta)$. The spectrum of the dipole has been confirmed to
be the differential of a blackbody spectrum \cite{microwave:fixsen94}.
At higher order there are
additional effects arising from aberration and from modulation of the
anisotropy pattern, which have also been observed \cite{microwave:Planck2013XXVII}.

The implied velocity for the Solar
System barycenter is $v = 369.82\pm 0.11\,{\rm km}\,{\rm s}^{-1}$, assuming a
value $T_0 = T_\gamma$, towards $(l,b) =
(264.021^{\circ}\pm0.011^{\circ},
48.253^{\circ}\pm0.005^{\circ} $) \cite{microwave:Planck2018I}.
This Solar System motion implies a velocity for the Galaxy and the Local
Group of galaxies relative to the CMB of
$v_{\rm LG} = 620 \pm 15\,{\rm km}\,{\rm s}^{-1}$ towards
$(l,b) = (271.9^{\circ} \pm 2.0^{\circ},
 29.6^{\circ} \pm 1.4^{\circ} $) \cite{microwave:Planck2018I};
most of the error comes from uncertainty in the velocity of the
Solar System relative to the Local Group.

The dipole is a frame-dependent quantity, and one can thus determine
the `CMB frame' (in some sense this is a special frame) as that in which the
CMB dipole would be zero.  Any velocity of the receiver relative to the Earth
and the Earth around the Sun is removed for the purposes of CMB anisotropy
studies, while our velocity relative to the Local Group of galaxies and
the Local Group's motion relative to the CMB frame are normally removed
for cosmological studies.  The dipole is now routinely used as a
primary calibrator for mapping experiments, either via the time-varying
orbital motion of the Earth, or through the cosmological
dipole measured by satellite experiments.

\subsection{Higher-Order Multipoles}
\label{microwave:sec:CMBmultipoles}
\index{CMB!multipoles}%

The variations in the CMB temperature maps
at higher multipoles ($\ell \geq 2$) are
interpreted as being mostly the result of density perturbations
in the early Universe, manifesting themselves at the epoch of the last
scattering of the CMB photons.  In the hot Big Bang picture, the expansion
of the Universe cools the plasma so that by
a redshift 
\index{Redshift!$z\simeq 1100$, reionization}%
$z\simeq1100$ (with little dependence on the
details of the model), the hydrogen and helium
nuclei can bind electrons into neutral atoms, a process usually referred to
as `recombination' \cite{microwave:SSS}.  Before this epoch, the CMB
photons were tightly coupled to the charged baryons, while afterwards they
could freely stream towards us.  By measuring the $a_{\ell m}$s we are thus
learning directly about physical conditions in the early Universe.

A statistically-isotropic sky means that all $m$s are equivalent, \ie,
there is no preferred axis, so that the temperature correlation function
between two positions on the sky
depends only on angular separation and not orientation.
Together with the assumption of Gaussian statistics (\ie,~no correlations
between the modes), the 2-point function of the
temperature field (or equivalently the power spectrum in $\ell$)
then fully characterizes the anisotropies.
The power summed over all $m$s at each $\ell$ is
$(2 \ell +1) C_\ell/(4\pi)$,
where $C_\ell \equiv \left\langle{|a_{\ell m}|^2}\right\rangle$.
Thus, averages of $a_{\ell m}$s over
$m$ can be used as estimators of the $C_\ell$s to constrain their
expectation values,
which are the quantities predicted by a theoretical model.  For an
idealized full-sky observation, the variance of each measured $C_\ell$
(\ie,~the variance of the variance) is $[2 /(2 \ell +1 )] C^2_\ell$.  This
sampling uncertainty (known as 
\index{CMB!cosmic variance}%
`cosmic variance') comes about because
each $C_\ell$ is $\chi^2$ distributed with $( 2 \ell +1 )$ degrees of
freedom for our observable volume of the Universe.  For fractional sky
coverage, $f_{\rm sky}$, this variance is increased by $1/f_{\rm sky}$
and the modes become partially correlated.

It is important to understand that theories predict the expectation
value of the power spectrum, whereas our sky is a {\it single\/} realization.
Hence, the cosmic variance is an unavoidable source of uncertainty
when constraining models; it dominates the scatter at lower $\ell$s,
while the effects of instrumental noise and resolution dominate at
higher $\ell$s \cite{Knox:1995dq}.

Theoretical models generally predict that the $a_{\ell m}$ modes are
Gaussian random fields to high precision, matching the empirical tests,
\eg,~standard
\index{Inflation!slow-roll}%
\index{Slow-roll inflation}%
slow-roll inflation's
non-Gaussian contribution is expected to be at least an order of
magnitude below current observational limits \cite{microwave:bartolo04}.
Although
\index{CMB!non-Gaussianity}%
\index{Non-Gaussianity CMB}%
non-Gaussianity of various forms is possible in early Universe models,
tests show that Gaussianity is an extremely good simplifying
approximation \cite{microwave:Planck2013XXIV}.  The only current
indications of any non-Gaussianity or statistical anisotropy are some
relatively weak signatures at large scales, seen in both
{\it WMAP\/} \cite{microwave:WMAPanomalies} and {\it Planck\/}
data \cite{microwave:PlanckIandS},
but not of high enough significance to reject the simplifying assumption.
Nevertheless, models that deviate from the inflationary slow-roll conditions
can have measurable non-Gaussian signatures.
So while the current observational limits make the power spectrum the
dominant probe of cosmology, it is worth noting that higher-order
correlations are tools for constraining otherwise viable theories.

\subsection{Angular Resolution and Binning}
\label{microwave:sec:CMBresolution}

There is no one-to-one conversion between multipole $\ell$ and
the angle subtended by a particular spatial scale projected onto the
sky.  However, crudely speaking, a single spherical harmonic $Y_{\ell m}$
corresponds to angular variations of $\theta\sim\pi/\ell$.  CMB maps contain
anisotropy information from the size of the map (or in practice some
fraction of that size) down to the beam-size of the instrument,
$\sigma$ (the standard deviation of the beam, in radians).
One can think of the effect of a Gaussian beam as rolling
off the power spectrum with the function ${\rm
e}^{-\ell(\ell+1)\sigma^2}$.

For less than full sky coverage, the $\ell$ modes become correlated.
Hence, experimental results are usually quoted as a series of `band
powers,' defined as estimators of $\ell(\ell+1)C_\ell/2\pi$ over
different ranges of $\ell$.  Because of the strong foreground signals
in the Galactic plane, even `all-sky' surveys, such as {\it WMAP\/} and
{\it Planck}, involve a cut sky.  The amount of binning required to
obtain uncorrelated estimates of power also depends on the map size.

\section{Cosmological Parameters}
\label{microwave:sec:CMBparameters}

The current `Standard Model' of cosmology contains around 10 free
parameters, only six of which 
are required to have non-null values (see The Cosmological
Parameters---Sec.~\ref{hubble} of the {\it Review of Particle Physics}).
The basic framework is the Friedmann-Robertson-Walker (FRW) metric (\ie,~a
Universe that is approximately homogeneous and isotropic on large
scales), with density perturbations laid down at early times and
evolving into today's structures (see Big-Bang
cosmology---Sec.~\ref{bigbang} of the {\it Review of Particle Physics}).
The most general possible set of density variations is a linear
combination of an adiabatic density perturbation and some isocurvature
perturbations.  Adiabatic means that there is no
change to the entropy per particle for each species,
\ie,~$\delta\rho/\rho$ for matter is $(3/4)\delta\rho/\rho$ for radiation.
Isocurvature means that the set of individual density perturbations adds\
to zero, for example, matter perturbations compensate radiation perturbations
so that the total
energy density remains unperturbed, \ie,~$\delta\rho$ for matter is
$-\delta\rho$ for radiation.  These different modes give rise to distinct
(temporal) phases during growth, with those of the adiabatic scenario being
fully consistent with the data.  Models that generate mainly isocurvature type
perturbations (such as most topological defect scenarios) are not
viable.  However, an admixture of the adiabatic mode with
up to 1.7\% isocurvature contribution (depending on details of the mode)
is still allowed \cite{microwave:Planck2018Infl}.

\subsection{Initial Condition Parameters}
\label{microwave:sec:CMBIC}

Within the adiabatic family of models, there is, in principle, a free
function describing the variation of comoving curvature perturbations,
${\cal R}({\bf x},t)$.  The great virtue of ${\cal R}$ is that
it is constant in time on super-horizon scales for a purely
adiabatic perturbation.  There are physical reasons to anticipate that the
variance of these perturbations will be described well by a power law
in scale, \ie,~in Fourier space
$\left\langle |{\cal R}|^2_k\right\rangle\propto k^{n_{\rm s}-4}$,
where $k$ is wavenumber and $n_{\rm s}$ is the spectral index as usually
defined.  So-called
`scale-invariant' initial conditions (meaning gravitational potential
fluctuations that are independent of $k$) correspond to $n_{\rm s}=1$.
In inflationary models \cite{microwave:lidlyth} (see
Inflation---Sec.~\ref{inflation} of the {\it Review of Particle Physics}),
perturbations are generated by
quantum fluctuations, which are set by the energy scale of inflation
together with the slope and higher derivatives of the inflationary potential.
One generally expects that the Taylor
series expansion of $\ln{\cal R}_k(\ln k)$ has terms of steadily decreasing
size.  For the simplest models, there are thus two parameters describing
the initial conditions for density perturbations, namely the amplitude and
slope of the power spectrum.  These can be explicitly defined, for
example, through
\begin{equation}
\label{microwave:eq:deltaR}
{\cal P}_{\cal R}^2\equiv k^3\left\langle |{\cal R}|_k^2\right\rangle
 /2\pi^2 \simeq A_{\rm s} \left(k/k_0\right)^{n_{\rm s}-1},
\end{equation}
with $A_{\rm s}\equiv{\cal P}_{\cal R}^2(k_0)$ and $k_0=0.05\,{\rm
Mpc}^{-1}$, say.  There are other equally valid definitions of the
amplitude parameter (see also Secs.~\ref{bigbang},
\ref{inflation}, and \ref{hubble}
of the {\it Review of Particle Physics}),
and we caution that the relationships between some of them can
be cosmology-dependent.  In slow-roll inflationary models, this
normalization is proportional to the combination $V^3/(V^\prime)^2$,
for the inflationary potential $V(\phi)$.  The slope $n_{\rm s}$ also involves
$V^{\prime\prime}$, and so the combination of $A_{\rm s}$ and $n_{\rm s}$ can
constrain potentials.

Inflation generates tensor (gravitational wave) modes, as well as scalar
(density perturbation) modes.  This property introduces another parameter,
measuring the amplitude of a possible tensor component, or equivalently the
ratio of the tensor to scalar contributions.  The tensor amplitude is
$A_{\rm t}\propto V$, and thus one expects a larger
gravitational wave contribution
in models where inflation happens at higher energies.
The tensor power spectrum also has a slope, often denoted $n_{\rm t}$,
but since this seems unlikely to be measured in the near future (and there is
also a consistency relation with tensor amplitude), it is sufficient
for now to focus only on the amplitude of the gravitational wave component.
It is most common to define the tensor contribution through $r$, the ratio
of tensor to scalar perturbation spectra
at some fixed value of $k$ (\eg,~$k=0.002\,{\rm Mpc}^{-1}$ or
$k=0.05\,{\rm Mpc}^{-1}$, although it was
historically defined in terms of the ratio of contributions at $\ell = 2$).
Different inflationary potentials will lead to different predictions,
\eg,~for 50 e-folds, $\lambda \phi^4$ inflation gives $r=0.32$ and
$m^2\phi^2$ inflation gives $r=0.16$ (both now strongly disfavored by the data),
while other models can have arbitrarily small values of $r$.
In any case, whatever the
specific definition, and whether they come from inflation or
something else, the `initial conditions' give rise to a minimum of
three parameters, $A_{\rm s}$, $n_{\rm s}$, and $r$.

\begin{figure}[htbp!]
\setlabel{pos=sw,fontsize=\scriptsize}
\xincludegraphics[width=0.99\columnwidth,label=\textsl{Scott \& Smoot 23}]{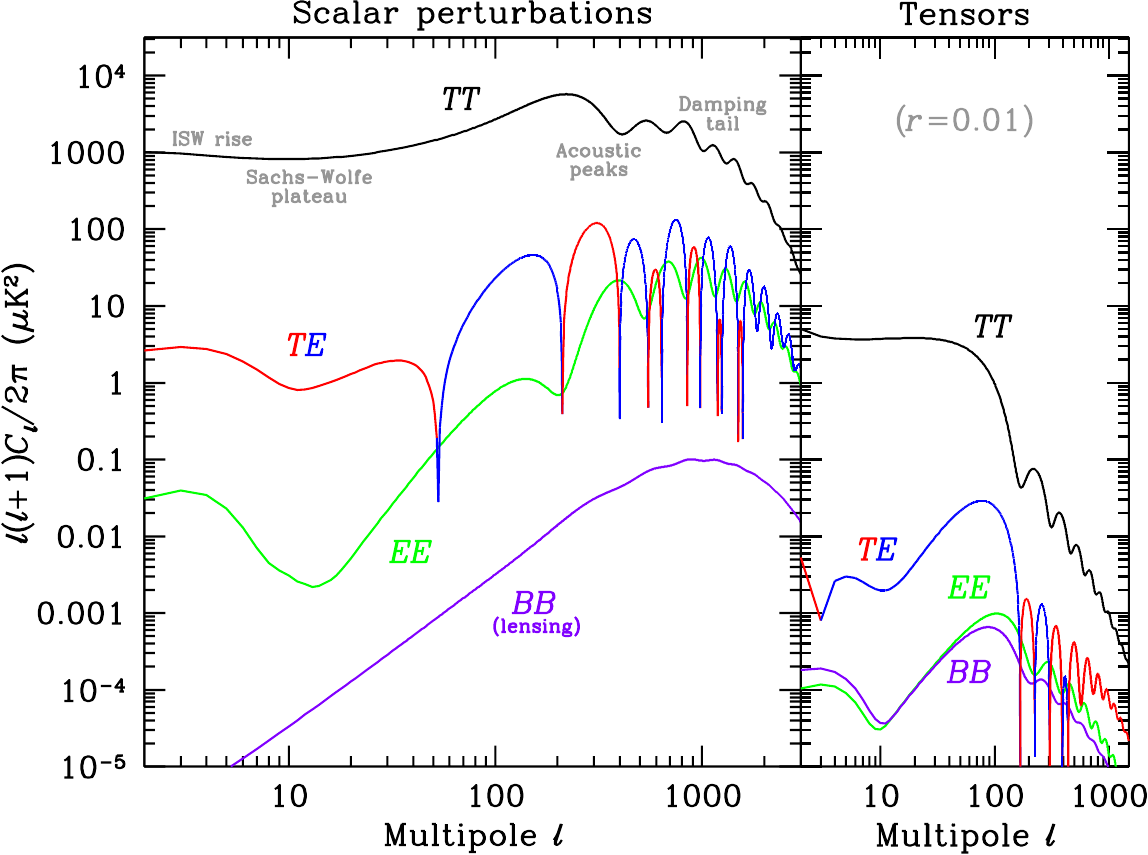}
\caption{Theoretical CMB anisotropy power spectra, using the best-fitting
${\rm \Lambda}$CDM model from {\it Planck}, calculated using {\tt CAMB}.
The panel on the left shows the theoretical expectation
for
\index{CMB!scalar perturbations}%
\index{CMB!tensor perturbations}%
scalar perturbations, while the panel on the right is for 
tensor perturbations, with an amplitude set to $r=0.01$ for illustration.
Note that the horizontal axis is logarithmic
here. For the well-measured scalar $TT$ spectrum, these regions, each covering
roughly a decade in $\ell$, are labeled as in the text: the ISW rise;
Sachs-Wolfe plateau; acoustic peaks; and damping tail.  The $TE$
cross-correlation power spectrum changes sign, indicated here by
plotting the absolute value, but switching color for the negative parts.}
\label{microwave:fig:CMBlabelled}
\end{figure}

\subsection{Background Cosmology Parameters}
\label{microwave:sec:CMBFRW}

\index{Background cosmology parameters}%
\index{Hubble!constant $H_0$}%
The FRW cosmology requires an expansion parameter (the Hubble
constant, $H_0$, often represented through
$H_0=100\,h\,{\rm km}\,{\rm s}^{-1}{\rm Mpc}^{-1}$)
and several parameters to describe the matter
and energy content of the Universe.  These are usually given in terms
of the critical density, \ie,~for species `x,'
${\rm \Omega}_{\rm x}\equiv\rho_{\rm x}/\rho_{\rm crit}$,
where $\rho_{\rm crit}\equiv3H_0^2/8\pi G$.  Since physical densities
$\rho_{\rm x}\propto{\rm \Omega}_{\rm x}h^2\equiv\omega_{\rm x}$
are what govern the
physics of the CMB anisotropies, it is these $\omega$s that are best
constrained by CMB data.  In particular, CMB observations constrain
${\rm \Omega}_{\rm b}h^2$ for baryons and ${\rm \Omega}_{\rm c}h^2$
for cold dark matter
(with $\rho_{\rm m}=\rho_{\rm c}+\rho_{\rm b}$ for the matter sum).

\index{Cosmological!constant ${\rm \Lambda}$}%
The contribution of a cosmological constant ${\rm \Lambda}$ (or other form
of dark energy, see Dark Energy---Sec.~\ref{darkenergy})
is usually included, together with a parameter that quantifies the curvature,
${\rm \Omega}_{K}\equiv 1-{\rm \Omega}_{\rm tot}$, where
${\rm \Omega}_{\rm tot}={\rm \Omega}_{\rm m}+{\rm \Omega}_{\rm \Lambda}$.
The radiation content,
while in principle a free parameter, is precisely enough determined by
the measurement of $T_\gamma$ that it can be considered fixed,
making a $<10^{-4}$ contribution to ${\rm \Omega}_{\rm tot}$ today.

Astrophysical processes at relatively low redshift can also affect the
$C_\ell$s, with a particularly significant influence during
reionization.  The Universe became reionized at some redshift $z_{\rm i}$,
long after recombination, affecting the CMB when it passed through the
integrated Thomson-scattering optical depth:
\begin{equation}
\label{microwave:eq:tau}
\tau = \int_0^{z_{\rm i}} \sigma_{\rm T} n_{\rm e}(z) {dt\over dz} dz,
\end{equation}
where $\sigma_{\rm T}$ is the Thomson cross-section, $n_{\rm e}(z)$ is the
number density of free electrons (which depends on astrophysics), and
$dt/dz$ is fixed by the background cosmology.  In principle, $\tau$
could be determined from the small-scale matter power spectrum, together with
the physics of structure formation and radiative feedback processes;
however, because this is a sufficiently intractable calculation, in practice
$\tau$ needs to be considered as a free parameter.

Thus, we have eight basic cosmological parameters, namely
$A_{\rm s}$, $n_{\rm s}$, $r$, $h$, ${\rm \Omega}_{\rm b}h^2$,
${\rm \Omega}_{\rm c}h^2$, ${\rm \Omega}_{\rm tot}$,
and $\tau$.  One can add additional parameters
to this list, particularly when using the CMB in combination with other data
sets.  The next most relevant ones might be: ${\rm \Omega}_\nu h^2$, the massive
neutrino contribution; $w$ ($\equiv p/\rho$), the equation of state
parameter for the dark energy;
and $dn_{\rm s}/d\ln k$, measuring deviations from a constant spectral index.
To these 11 one could of course add further parameters describing additional
physics, such as details of the reionization process, features
in the initial power spectrum, a sub-dominant contribution of isocurvature
modes, {\it etc}.

As well as these underlying parameters, there are other (dependent) quantities
that can be obtained from them.  Such derived parameters include the actual
${\rm \Omega}$s of the various components (\eg,~${\rm \Omega}_{\rm m}$),
the variance of density perturbations at particular scales (\eg,~$\sigma_8$),
the angular scale of the sound horizon ($\theta_\ast$),
the age of the Universe today ($t_0$),  the age of the Universe at
recombination, reionization, {\it etc.} (see The Cosmological
Parameters---Sec.~\ref{hubble}).

\section{Physics of Anisotropies}
\label{microwave:sec:CMBphysics}

The cosmological parameters affect the anisotropies through the well
understood physics of the evolution of linear perturbations within a
background FRW cosmology.
There are very effective, fast, and publicly-available software codes
for computing the CMB temperature, polarization, and matter power spectra,
\eg,~{\tt CMBFAST} \cite{microwave:SelZal},
{\tt CAMB} \cite{microwave:LewChaLas}, and {\tt CLASS} \cite{microwave:class}.
These have been tested over a wide range of cosmologies and are
considered to be accurate to much better than the 1\%
level \cite{Seljak:2003th}, so that numerical errors are less than
10\% of the parameter uncertainties for
{\it Planck\/} \cite{microwave:PlanckParams}.

For pedagogical purposes, it is easiest to focus on the temperature
anisotropies, before moving to the polarization power spectra.
A description of the physics underlying the $C^{TT}_\ell$s can be separated
into four main regions (the first two combined below),
as shown in the top left part of Fig.~\ref{microwave:fig:CMBlabelled}.

\subsection{The ISW Rise, $\ell\,{\lesssim}\,10$, and Sachs-Wolfe Plateau,
$10\,{\lesssim}\,\ell\,{\lesssim}\,100$}
\label{microwave:sec:CMBlargel}

\index{Horizon scale}%
\index{Hubble!radius}%
The horizon scale (or more precisely, the angle subtended by the Hubble
radius) at last scattering corresponds to $\ell\,{\simeq}\,100$.
Anisotropies at larger scales have not evolved significantly, and hence
directly reflect the `initial conditions.'  Temperature variations are
$\delta T/T = -(1/5)  {\cal R}({\bf x_{\rm LSS}}) \simeq (1/3) \delta\phi/c^2$,
where $\delta\phi$ is the
perturbation to the gravitational potential, evaluated on the 
\index{Last scattering surface, LSS}%
last-scattering surface (LSS).  This is a result of the
combination of gravitational redshift and intrinsic temperature fluctuations,
and is usually referred to as the Sachs-Wolfe effect \cite{microwave:SacWol}.
\index{Sachs-Wolfe effect}%

Assuming that a nearly scale-invariant spectrum of curvature
(and corresponding density) perturbations was laid down at early times
(\ie,~$n_{\rm s}\simeq1$, meaning equal power per decade
in $k$), then $\ell(\ell+1)C_\ell \simeq {\rm constant}$
at low $\ell$s.  This predicted near-flatness is hard to see unless the
multipole axis is plotted logarithmically (as in
Fig.~\ref{microwave:fig:CMBlabelled},
and part of Fig.~\ref{microwave:fig:Cldata}).

Time variation of the potentials (\ie,~time-dependent metric perturbations)
at late times
leads to an upturn in the $C_\ell$s in the lowest several multipoles;
any deviation from a total equation of state $w=0$ has such an effect.
So the dominance of the dark energy at low redshift
(see Dark Energy---Sec.~\ref{darkenergy})
makes the lowest $\ell$s rise above the plateau.  This is usually
called
\index{Integrated Sachs-Wolfe effect}%
\index{Sachs-Wolfe effect, integrated}%
the integrated Sachs-Wolfe effect (or ISW rise),
since it comes from the line integral of~${\dot \phi}$;
it has been confirmed through correlations between the large-angle
anisotropies and large-scale
structure \cite{microwave:iswcorr,*microwave:Planck2015ISW}.
Specific models can also give additional contributions at low~$\ell$
(\eg,~perturbations in the dark-energy component
itself \cite{microwave:HETW}), but
typically these are buried in the cosmic variance.

\index{CMB!primordial perturbation}%
In principle, the mechanism that produces primordial perturbations could
generate scalar, vector, and tensor modes.  However, the vector (vorticity)
modes decay with the expansion of the Universe.  The tensors (transverse
trace-free perturbations to the metric) generate temperature anisotropies
through the integrated effect of the locally-anisotropic expansion of
space.  Since the tensor modes also redshift away after
they enter the horizon, they contribute only to angular scales above
about $1^\circ$ (see Fig.~\ref{microwave:fig:CMBlabelled}).
Hence, some fraction of the low-$\ell$ signal could be due to a gravitational
wave contribution, although small amounts of tensors are essentially
impossible to discriminate from other effects that might raise the level
of the plateau.  Nevertheless, the tensors {\it can\/} be distinguished
using polarization information (see Sec.~\ref{microwave:sec:CMBpolarization}).

\subsection{The Acoustic Peaks, $100\lesssim\ell\lesssim1000$}
\label{microwave:sec:CMBmidl}

On sub-degree scales, the rich structure in the anisotropy
spectrum is the consequence of gravity-driven acoustic oscillations occurring
before the atoms in the Universe became neutral \cite{microwave:HuSugiyamaSilk}.
Perturbations inside the horizon at last scattering were able to evolve
causally and produce anisotropy at the last-scattering epoch, which reflects
this evolution.
The frozen-in phases of these sound waves imprint a dependence
on the cosmological parameters, which gives CMB anisotropies their great
constraining power.

The underlying physics can be understood as follows.  Before the Universe
became neutral, the
proton-electron plasma was tightly coupled to the photons, and these
components behaved as a single `photon-baryon fluid.'
Perturbations in the gravitational potential, dominated by the dark-matter
component, were steadily evolving.  They drove oscillations in
the photon-baryon fluid, with photon pressure providing most of the restoring
force and baryons giving some additional inertia.
The perturbations were quite small in amplitude, ${\cal O}(10^{-5})$, and so
evolved linearly. That means each Fourier mode developed independently, and
hence can be described as a driven harmonic oscillator, with frequency
determined by the sound speed in the fluid.  Thus, the fluid density
underwent oscillations,
giving time variations in temperature.  These combine with a
velocity effect, which is $\pi/2$ out of phase and has its amplitude
reduced by the sound speed.

After the Universe recombined, the radiation decoupled from the baryons and
could travel freely towards us.  At that point, the (temporal) phases
of the oscillations were frozen-in, and became projected on the sky as a
harmonic series of peaks and troughs in power.
The main peak is the mode that went through
1/4 of a period, reaching maximal compression.  The even peaks are
maximal {\it under\/}-densities, which are generally of smaller
amplitude because the rebound has to fight against the baryon inertia.
The troughs, which do not extend to zero power, are partially filled by
the Doppler effect because they are at the velocity maxima.

The physical length scale
associated with the peaks is the sound horizon at last scattering,
which can be straightforwardly calculated.  This
length is projected onto the sky, leading to an angular scale that
depends on the geometry of space, as well as the distance to last
scattering.  Hence, the angular position of the
peaks is a sensitive probe of a particular combination of cosmological
parameters.  In fact, this characteristic angular scale, $\theta_\ast$, is the
most precisely measured observable, and hence is usually treated as an element
of the cosmological parameter set.

One additional effect arises from reionization at redshift $z_{\rm i}$.  A
fraction of photons ($\tau$) will be isotropically scattered at
$z\,{<}\,z_{\rm i}$,
partially erasing the anisotropies at angular scales smaller than those
subtended by the Hubble radius at $z_{\rm i}$.  This corresponds
typically to $\ell$s above about 10, depending on the specific
reionization model.  The acoustic peaks are therefore reduced by a
factor $e^{-2\tau}$ relative to the plateau.

These acoustic peaks were a clear theoretical prediction going back to
about 1970 \cite{microwave:PeeYu,*microwave:SunZel70}.
One can think of them as a snapshot of
stochastic standing waves.  Since the physics governing them is
simple and their structure rich, one can see how they encode extractable
information about the cosmological parameters.
Their empirical existence started to become
clear around 1994 \cite{microwave:scott95}, and the emergence,
over the following
decade, of a coherent series of acoustic peaks and troughs is a triumph
of modern cosmology.  This picture has received further confirmation with the
detection in the power spectrum of galaxies (at redshifts $z\,{\lesssim}\,1$)
of the imprint of these same acoustic oscillations in the baryon
component \cite{microwave:eisenstein05},
as well as through detection of the expected oscillations in CMB polarization
power spectra (see Sec.~\ref{microwave:sec:CMBpolarization}).

\subsection{The Damping Tail, $\ell\gtrsim1000$}
\label{microwave:sec:CMBhighl}

The recombination process is not instantaneous, and this imparts a thickness to
the LSS.  This leads to a damping of the anisotropies at
the highest $\ell$s, corresponding to scales smaller than that subtended
by this thickness.  One can also think of the photon-baryon fluid as having
imperfect coupling, so that there is diffusion between the two components,
and hence the amplitudes of the oscillations decrease with time.
These effects lead to a damping of
the $C_\ell$s, sometimes called 
\index{Silk damping}%
`Silk damping' \cite{microwave:silk},
which cuts off the anisotropies at multipoles above about 2000.  So, although
in principle it is possible to measure to ever smaller scales, this becomes
increasingly difficult in practice.

\subsection{Gravitational Lensing Effects}
\label{microwave:sec:CMBlensing}
\index{Gravitational lensing}%

CMB gravitational lensing is caused by
structures at lower redshift along the line of sight to the LSS.  Photon paths
are deflected by the lensing potential $\phi$, such that
$T({\hat n})\rightarrow T({\hat n}+\nabla\phi)$.  Typical deflections are
around $2\,{\rm arcmin}$, but coherent over scales of a degree or so.  Lensing
preserves surface brightness, which means that a uniform temperature field
would be unaffected; however, since there are anisotropies, then several
distinct effects can be identified.  The $C_\ell$s are
convolved with a smoothing function in a calculable way, partially flattening
the peaks and troughs, generating a power-law tail at the highest multipoles,
and complicating the polarization signal \cite{Zaldarriaga:1998ar} (see
Sec.~\ref{microwave:sec:CMBBB}).
Additionally, the effect of lensing on the CMB can be detected
through the 4-point function, which correlates
temperature gradients and small-scale anisotropies, enabling a map of the
lensing potential to be constructed \cite{microwave:PlanckLens},

Lensing is important because it gives an independent estimate of
$A_{\rm s}$, breaking the parameter combination $A_{\rm s}{\rm e}^{-2\tau}$
that is largely degenerate in the temperature anisotropy power spectra.
Lensing is an example of a `secondary effect,'
\ie,~the processing of anisotropies due to relatively nearby
structures (see Sec.~\ref{microwave:sec:CMBsecondaries}).
Galaxies and clusters of galaxies
give several such effects; all are expected to be of low amplitude, but are
increasingly important at the highest $\ell$s.  Such effects carry additional
cosmological information (about evolving gravitational potentials in the
low-redshift Universe) and are receiving more attention as experiments
push to higher sensitivity and angular resolution.  The lensing power spectrum
(see Sec.~\ref{microwave:sec:CMBphiphi})
can potentially constrain dark-energy evolution, while future measurements
at high $\ell$ are a particularly sensitive
probe of the sum of the neutrino masses \cite{microwave:kaplighat03}.

\begin{figure}[htbp!]
\setlabel{pos=sw,fontsize=\scriptsize}
\xincludegraphics[width=0.99\columnwidth,label=\textsl{Scott \& Smoot 23}]{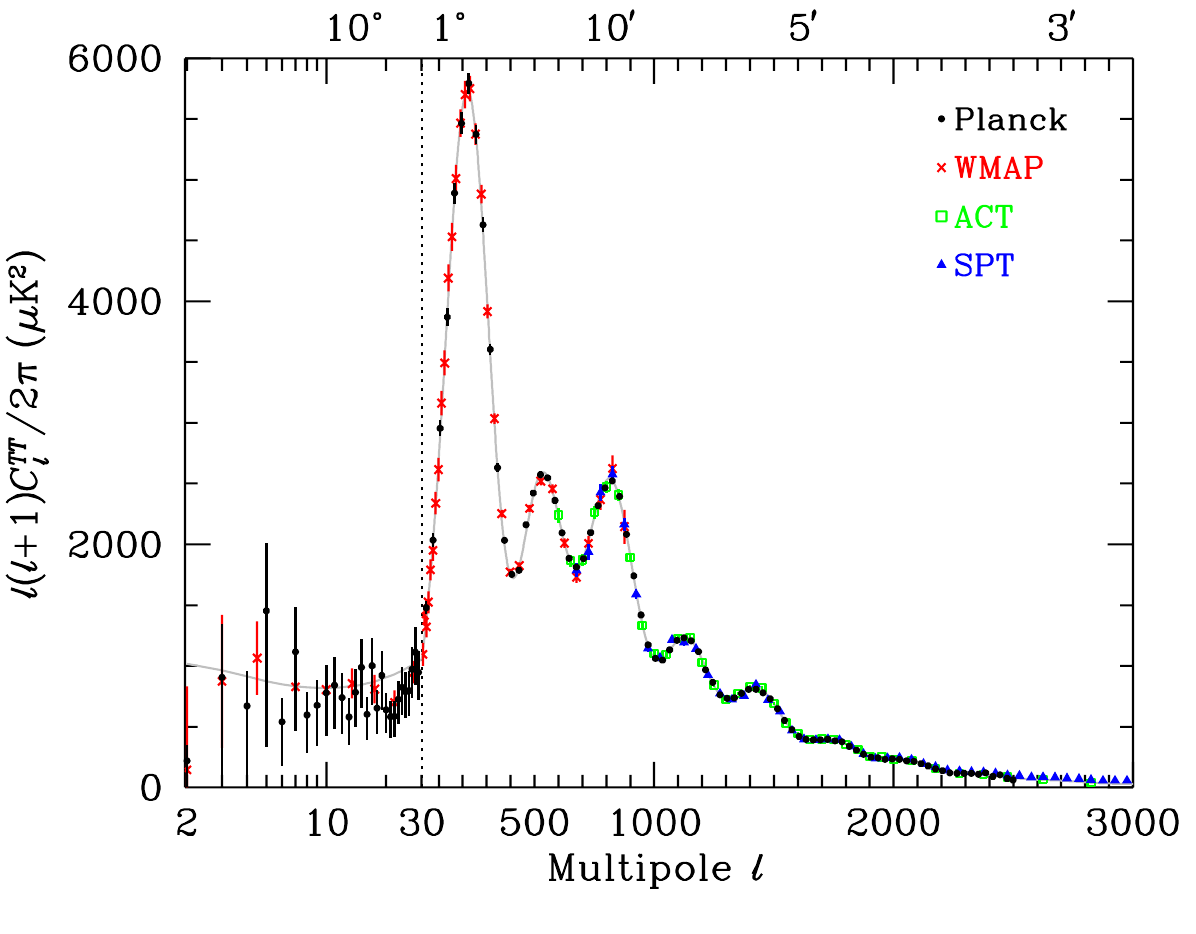}
\caption{CMB temperature anisotropy
band-power estimates from the {\it Planck}, {\it WMAP}, ACT, and SPT
experiments.  Note that the widths of the multipole bands vary between
experiments and have not been plotted.
This figure represents only a selection of the most recent
available experimental results,
and some points with large error bars have been omitted.
At the higher multipoles these band-powers involve
subtraction of particular foreground models, and so proper analysis
requires simultaneous fitting of CMB and foregrounds over multiple frequencies.
The horizontal axis here is logarithmic for the lowest multipoles,
to show the Sachs-Wolfe plateau, and linear for the other multipoles.
The acoustic peaks and damping region are very clearly
observed, with no need for a theoretical line to guide the eye; however,
the curve plotted is the best-fit {\it Planck\/} ${\rm \Lambda}$CDM model.}
\label{microwave:fig:Cldata}
\end{figure}

\section{Current Temperature Anisotropy Data}
\label{microwave:sec:CMBdata}

Steady improvement in CMB data quality has led to the present-day cosmological
model.  The most robust constraints currently available come
from
\index{Satellite!{\it Planck}, CMB}%
\index{CMB!{\it Planck} satellite}%
{\it Planck\/}
satellite \cite{microwave:PlanckLike,microwave:Planck2018Like} data
(together with constraints from non-CMB cosmological data sets),
although smaller-scale results from the ACT \cite{microwave:das13} and
SPT \cite{microwave:story13} experiments are beginning to add useful
constraining power.
We plot power spectrum estimates from these experiments in
Fig.~\ref{microwave:fig:Cldata},
along with {\it WMAP\/} data \cite{microwave:hinshaw12} for comparison
(see previous versions of this review for data from earlier
experiments).  Independent experimental data show
consistency, both in maps and in derived power
spectra (up to systematic uncertainties in the overall calibration for
some experiments).  This makes it clear that systematic effects are largely
under control.

The band-powers shown in Fig.~\ref{microwave:fig:Cldata}
are in very good agreement with a `${\rm \Lambda}$CDM' model.  As described
earlier, several (at least seven) of the peaks and troughs are quite apparent.
The original papers present the details on how these estimates were made, on
the band-power correlation strengths, and on the information needed for their
proper interpretation.

\section{CMB Polarization}
\label{microwave:sec:CMBpolarization}
\index{CMB!polarization}%

Thomson scattering of an anisotropic radiation field also generates
linear polarization and the CMB is predicted to be polarized, at the level of
roughly 5\% of the temperature anisotropies \cite{microwave:hu97}.
Polarization is a spin-2 field on the sky, and the algebra of the modes
in multipole space is strongly analogous to spin-orbit coupling in quantum
mechanics \cite{Hu:1997hp}.
The linear polarization pattern can be decomposed in a number
of ways, with two quantities required for each pixel in a map, often given as
the $Q$ and $U$ Stokes parameters.  However, the most
intuitive and physical decomposition is a geometrical one, splitting
the polarization
pattern into a part that comes from a divergence (often referred to as
the `$E$ mode') and a part with a curl
(called the `$B$ mode') \cite{Zaldarriaga:1996xe,*microwave:KKSpol}.
More explicitly, the modes are defined in terms of second derivatives of
the polarization amplitude, with the Hessian for the $E$ modes having principal
axes in the same sense as the polarization, while the $B$-mode pattern can
be thought of as a $45^\circ$ rotation of the $E$-mode pattern.
Globally one sees that the $E$ modes have $(-1)^\ell$ parity (like the
spherical harmonics), while the $B$ modes have $(-1)^{\ell+1}$ parity.

The existence of this linear polarization allows for six different cross-power
spectra to be determined from data that measure the full temperature
and polarization anisotropy information.
Parity considerations make two of these zero, and we are
left with four potential observables, $C^{TT}_\ell$, $C^{TE}_\ell$,
$C^{EE}_\ell$, and $C^{BB}_\ell$ (see Fig.~\ref{microwave:fig:CMBlabelled}).
Because scalar perturbations have no handedness,
the $B$-mode power spectrum can only be sourced by vectors or tensors.
Moreover, inflationary scalar perturbations give only $E$ modes,
while tensors generate roughly equal amounts of $E$ and $B$,
therefore the determination of a non-zero $B$-mode signal is a way to measure
the gravitational-wave contribution (and thus potentially derive the energy
scale of inflation).  However, because the signal is expected to be
rather weak, one must first eliminate the foreground contributions
and other systematic effects down to very low levels.
In addition, CMB lensing creates $B$ modes from $E$ modes, further
complicating the extraction of a tensor signal.

As with temperature,
the polarization $C_\ell$s exhibit a series of acoustic peaks
generated by the oscillating photon-baryon fluid.
The main `$EE$' power spectrum has
peaks that are out of phase with those in the `$TT$' spectrum because the
polarization anisotropies are sourced by the fluid velocity.  The `$TE$'
part of the polarization and temperature patterns comes from
correlations between density and velocity perturbations on the last-scattering
surface, which can be both positive and negative, and is of larger amplitude
than the $EE$ signal.  There is no polarization Sachs-Wolfe effect,
and hence no large-angle (low-$\ell$) plateau.  However, scattering
during a recent period of reionization can create a polarization `bump'
at large angular scales.

\begin{figure}[htbp!]
\setlabel{pos=sw,fontsize=\scriptsize}
\xincludegraphics[width=0.99\columnwidth,label=\textsl{Scott \& Smoot 23}]{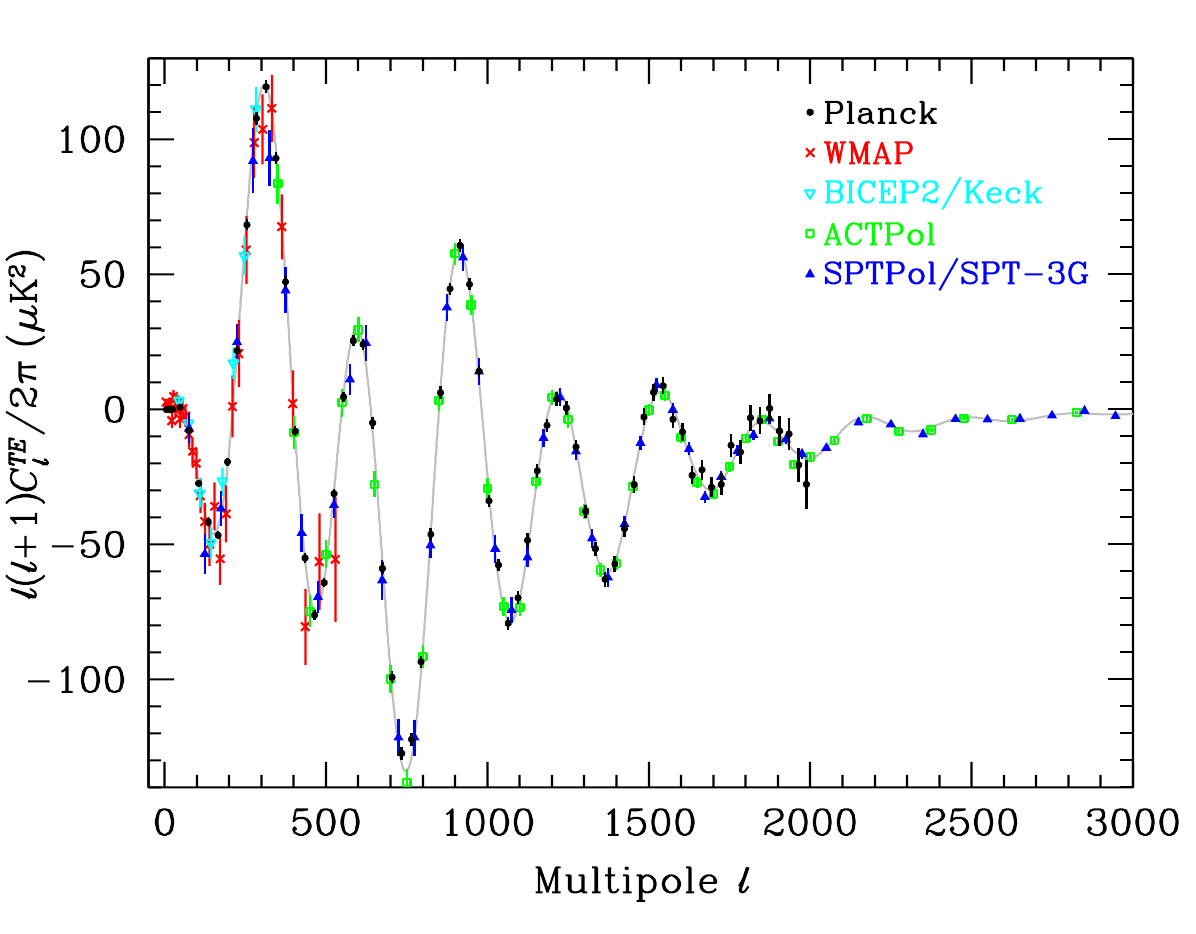}
\caption{Cross-power spectrum band-powers of the temperature anisotropies and
$E$-mode polarization signal from {\it Planck\/} (the low multipole data have
been binned here), as well as {\it WMAP},
BICEP2/Keck, ACTPol, and SPTPol/SPT-3G.  The curve is the best fit to the
{\it Planck\/} temperature, polarization, and lensing band-powers.
Note that each data point is an average over
a band of multipoles, and hence to compare in detail with a model one has
to integrate the theoretical curve through the band.}
\label{microwave:fig:CLTE}
\end{figure}

Because the polarization anisotropies have only a small fraction of the
amplitude of the temperature anisotropies, they took longer to detect.
The first measurement of a polarization signal came in 2002 from the DASI
experiment \cite{microwave:dasipol},
which provided a convincing detection, confirming the general paradigm,
but of low enough significance that it lent no real
constraint to models.  Despite dramatic progress since then,
it is still the case that polarization data mainly
support the basic paradigm, while reducing error bars on
parameters by only around 20\%.  However, there are exceptions to this,
specifically in the reionization optical depth, and the potential to
constrain primordial gravitational waves.  Moreover, the situation is
expected to change dramatically as more of the available polarization modes are
measured.

\subsection{\textit{T--E} Power Spectrum}
\label{microwave:sec:CMBTE}

Because the $T$ and $E$ skies are correlated, one has to measure the $TE$
power spectrum, as well as $TT$ and $EE$, in order to extract all the
cosmological information.
This $TE$ signal has now been mapped out extremely accurately by
{\it Planck\/} \cite{microwave:Planck2018Like}, and these band-powers
are shown in Fig.~\ref{microwave:fig:CLTE}, along with those from
{\it WMAP\/} \cite{microwave:larson11}
and BICEP2/Keck \cite{microwave:BKV}, with
ACTPol \cite{microwave:Choi20}
and SPTPol/-SPT-3G \cite{microwave:SPTPol,microwave:Balkenhol2023}
extending to smaller angular scales.
The anti-correlation at $\ell\simeq150$ and the peak at $\ell\simeq300$ were
the first features to become distinct, but now a whole series of oscillations
is clearly seen in this power spectrum (including six or seven peaks and
troughs \cite{microwave:Planck2018I}).
The measured shape of the cross-correlation power spectrum provides
supporting evidence for the general cosmological picture, and also
directly constrains the thickness of the last-scattering surface.
Because the polarization anisotropies are generated in this scattering
surface, the existence of correlations at angles above about a degree
demonstrates that there were super-Hubble fluctuations at the recombination
epoch.  The sign of this correlation also confirms the adiabatic paradigm.

The overall picture of the source of CMB polarization and its oscillations
has also been confirmed through tests that average the
maps around both temperature hot spots and cold
spots \cite{microwave:Planck2018IandS}.  One sees precisely the expected
patterns of radial and tangential polarization configurations, as well as the
phase shift between polarization and temperature.  This leaves no doubt that
the oscillation picture is the correct one and that the polarization is
coming from Thomson scattering at $z\simeq1100$.

\begin{figure}[htbp!]
\setlabel{pos=sw,fontsize=\scriptsize}
\xincludegraphics[width=0.99\columnwidth,label=\textsl{Scott \& Smoot 23}]{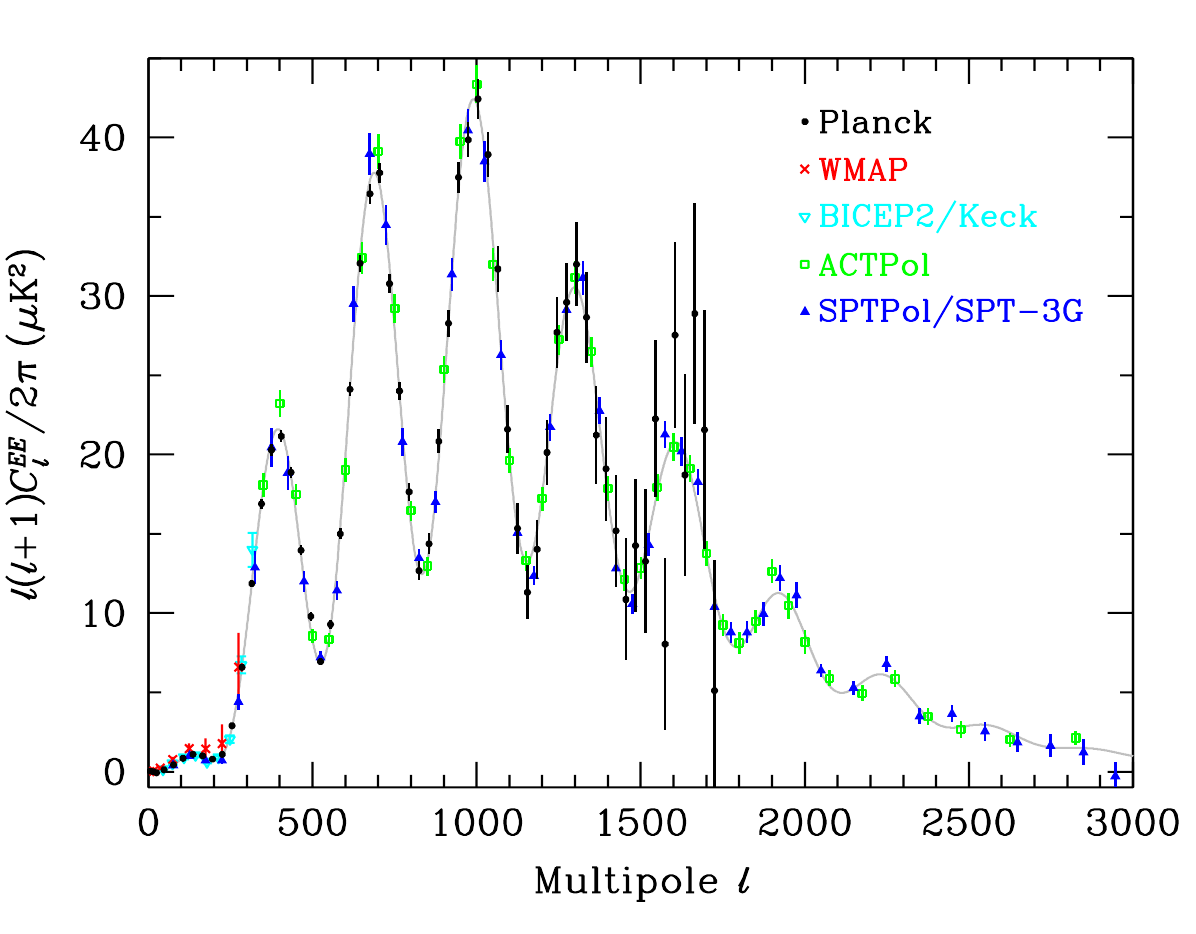}
\caption{Power spectrum of $E$-mode polarization from {\it Planck},
together with {\it WMAP}, BICEP2/Keck, ACTPol, and SPTPol/SPT-3G.
Note that some band-powers with
larger uncertainties have been omitted and that the unbinned {\it Planck\/}
low-$\ell$ data have been binned here.  Also
plotted is the best-fit theoretical model from {\it Planck\/} temperature +
polarization + lensing data.}
\label{microwave:fig:CLEE}
\end{figure}

\subsection{\textit{E--E} Power Spectrum}
\label{microwave:sec:CMBEE}

Experimental band-powers for $C_\ell^{EE}$ from {\it Planck}, {\it WMAP},
BICEP2/Keck Array \cite{microwave:BKV},
ACTPol \cite{microwave:Choi20},
and SPTPol/SPT-3G \cite{microwave:SPTPol,microwave:Balkenhol2023}
are shown in Fig.~\ref{microwave:fig:CLEE}.  Without the
benefit of correlating with the temperature anisotropies (\ie,~measuring
$C_\ell^{TE}$), the polarization
anisotropies are very weak and challenging to measure.  Nevertheless, the
oscillatory pattern is now well established and the data closely match the
$TT$-derived theoretical prediction.  In Fig.~\ref{microwave:fig:CLEE} one can clearly
see the `shoulder' expected at $\ell\simeq140$, the first main peak at
$\ell\simeq400$ (corresponding to the first trough in $C_\ell^{TT}$),
and the series of oscillations that is out of phase with those of the
temperature anisotropy power spectrum (including six or seven peaks and
troughs \cite{microwave:Planck2018I}).

Perhaps the most unique result from the polarization measurements
is at the largest angular scales ($\ell<10$) in $C_\ell^{TE}$ and
$C_\ell^{EE}$, where there is evidence for an excess signal
(not visible in Fig.~\ref{microwave:fig:CLEE}) compared to
that expected from the temperature power spectrum alone.  This is precisely
the signal anticipated from an early period of reionization, arising from
Doppler shifts during the partial scattering at $z<z_{\rm i}$.
The amplitude of the signal
indicates that the first stars, presumably the source of the
ionizing radiation, formed around $z\simeq8$ (although the uncertainty is
still quite large).  This corresponds to a scattering optical depth
$\tau\simeq0.06$, so roughly 6\% of CMB photons were re-scattered at the
reionization epoch, with the other 94\% last scattering at $z\simeq1100$.
However, estimates of the amplitude of this reionization excess have
come down since the
first measurements by {\it WMAP\/} (indicating that this is an extremely
difficult measurement to make) and the latest {\it Planck\/} results have
reduced the value further \cite{microwave:Planck2018Params,microwave:Tristram21}.

\begin{figure}[htbp!]
\setlabel{pos=sw,fontsize=\scriptsize}
\xincludegraphics[width=0.99\columnwidth,label=\textsl{Scott \& Smoot 23}]{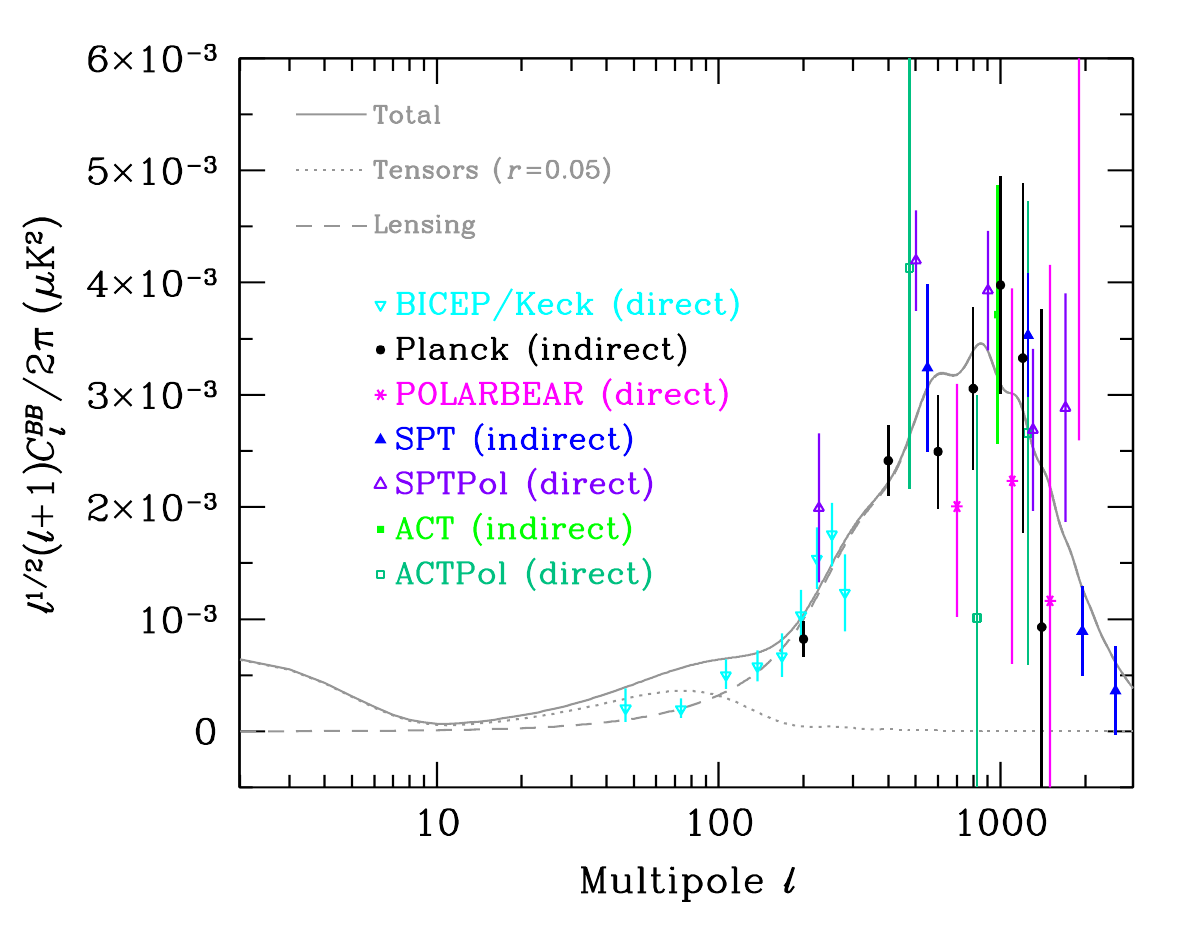}
\caption{Power spectrum of $B$-mode polarization, including results from the
BICEP2/BICEP3/Keck Array combined analysis, {\it Planck},
POLARBEAR, SPT, and ACT.  Note that some of the measurements are direct
estimates of $B$ modes on the sky, while others are only sensitive to the
lensing signal and come from combining $E$-mode and lensing potential
measurements.  Several earlier experiments
reported upper limits, which are all off the top of this plot.
A logarithmic horizontal axis is adopted here and the $y$-axis has been
divided by a factor of $\sqrt\ell$ in order to show all three theoretically
expected contributions: the low-$\ell$ reionization bump; the $\ell\simeq100$
recombination peak; and the high-$\ell$ lensing signature.
The dotted line is for a tensor (primordial gravitational wave) fraction
$r\,{=}\,0.05$, simply as an example, with all other cosmological parameters
set at the best {\it Planck}-derived values, for which model the expected
lensing $B$ modes have also been shown with a dashed line.}
\label{microwave:fig:CLBB}
\end{figure}

\subsection{\textit{B--B} Power Spectrum}
\label{microwave:sec:CMBBB}

The expected amplitude of $C_\ell^{BB}$ is very small, so measurements
of this polarization curl-mode are extremely challenging.
The first indication of
the existence of the $BB$ signal came from the detection of the
expected conversion of $E$ modes to $B$ modes by gravitational
lensing, through a correlation technique using the lensing potential and
polarization measurements from SPT \cite{microwave:Hanson13}.
However, the real promise of $B$ modes lies in the detection of
primordial gravitational waves at larger scales.  This tensor signature
could be seen either in the `recombination bump' at around $\ell=100$ (caused
by an ISW effect as gravitational waves redshift away at the
last-scattering epoch) or the `reionization bump' at $\ell\lesssim10$
(from additional scattering at low redshifts).

Results from the BICEP2 experiment \cite{microwave:BICEP2} in 2014
suggested a detection of the primordial $B$-mode signature around the
recombination peak.  BICEP2 mapped a small part of the CMB sky with the
best sensitivity level reached at that time (below
$100\,$nK), but at a single frequency.  Higher frequency data from
{\it Planck\/} indicated that much of the BICEP2 signal was due to dust
within our Galaxy, and a combined analysis by the BICEP2, Keck Array, and
{\it Planck\/} teams \cite{microwave:BKP} indicated that the data are
consistent with no primordial $B$ modes.  The current constraint from
{\it Planck\/} data alone is $r<0.069$ (95\% at
$k=0.05\,{\rm Mpc}^{-1}$\cite{microwave:Planck2018Params,microwave:Tristram21})
using all CMB power spectra,
and this limit is reduced to $r<0.044$ with the inclusion
of BICEP2/Keck Array data \cite{microwave:BK18,microwave:Tristram21}.
The most constraining limit is $r<0.036$ from a combination of BICEP2, Keck
Array, and BICEP3 data, using {\it WMAP\/} and {\it Planck\/} maps to help
remove foregrounds \cite{microwave:BKXIII}.

Several experiments are continuing to push down the sensitivity of $B$-mode
measurements, motivated by the enormous importance of a future detection
of this telltale signature of inflation (or other physics at the
highest energies).  A compilation of experimental results for $C_\ell^{BB}$
is shown in Fig.~\ref{microwave:fig:CLBB}, coming from a combination of direct estimates
of the $B$ modes (BICEP2/BICEP3/Keck Array \cite{microwave:BKXIII},
POLARBEAR \cite{microwave:POLARBEAR},
SPTPol \cite{microwave:Sayre2020}, and ACTPol \cite{microwave:Choi20})
and indirect determinations of the lensing $B$ modes based on estimating the
effect of measured lensing on measured $E$ modes
({\it Planck\/} \cite{microwave:Planck2015Lens},
SPT \cite{microwave:Hanson13},
and ACT \cite{microwave:vanEngelen15}).
Additional band-power estimates are expected from these and
other experiments in the near future, with the Simons
Observatory \cite{microwave:SimonsO}, the so-called `Stage 4'
CMB project \cite{microwave:CMBS4} and the
\index{Satellite!LiteBIRD, inflation, CMB}%
\index{CMB!LiteBIRD satellite}%
{\it LiteBIRD\/} satellite \cite{microwave:LiteBIRD23},
holding great promise for pushing down to the $r\sim0.001$ level.

\section{CMB Lensing Power Spectrum}
\label{microwave:sec:CMBphiphi}
\index{Gravitational lensing}%

\begin{figure}[htbp!]
\setlabel{pos=sw,fontsize=\scriptsize}
\xincludegraphics[width=0.99\columnwidth,label=\textsl{Scott \& Smoot 23}]{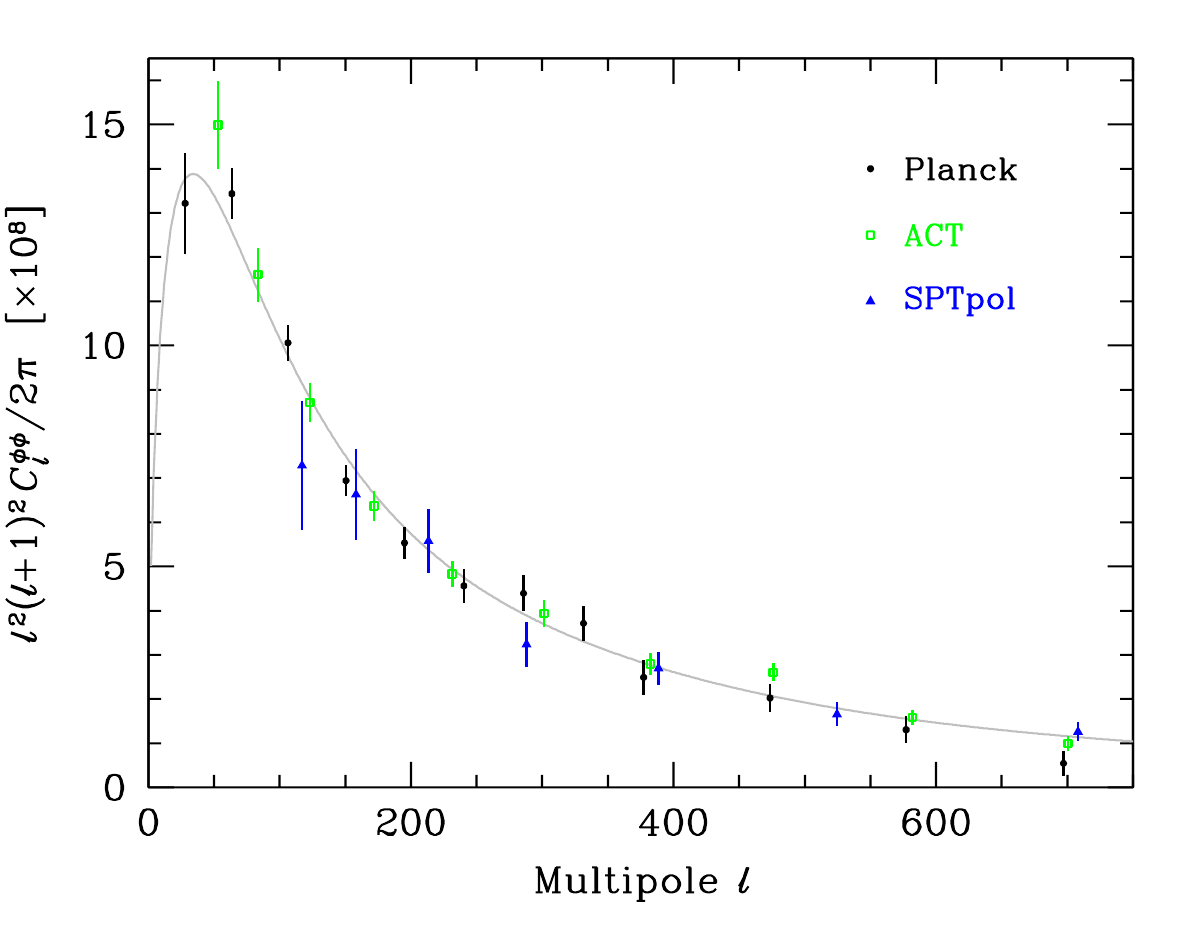}
\caption{Power spectrum measurements for CMB lensing, including selected
results from {\it Planck}, ACT, and SPT.  The quantity plotted is the
(dimensionless) potential power spectrum, scaled by $\ell^2(\ell+1)^2$ and
multiplied by a factor of $10^8$ to make the numbers more manageable.
Some less reliable measurements have not been plotted.  The best-fit
{\it Planck\/} $\Lambda$CDM spectrum is plotted as a gray line.}
\label{microwave:fig:CLphiphi}
\end{figure}

One further CMB observable is the gravitational lensing
deflection, leading to the construction of a map of the lensing potential.
The latest {\it Planck\/} results \cite{microwave:Planck2018Lens,*microwaveCarron2022} give a
map that is detected at the ${>}\,40\sigma$ level using a minimum-variance
procedure from the 4-point function of temperature and polarization data.
From these maps, estimates can be constructed of $C_\ell^{\phi\phi}$, the
lensing-potential power spectrum, which is found to be consistent with
predictions from the best-fit temperature and polarization model.
Recent results from ACT give a power spectrum that has a similar overall
signal-to-noise ratio \cite{microwave:Qu2023} and there are also interesting
measurements from SPT \cite{microwave:Wu2019}.
Figure~\ref{microwave:fig:CLphiphi} plots the {\it Planck}, ACT, and SPT
estimates of $C_\ell^{\kappa\kappa}$, the lensing convergence
power spectrum, which is proportional to $\ell^2(\ell+1)^2$ times
the potential power spectrum $C_\ell^{\phi\phi}$.

We can think of each sky pixel as possessing three independent quantities
that can be measured, namely $T$, $E$, and $\phi$ (and potentially $B$,
if that becomes detectable).  Determining the constraining power comes
down to counting $Y_{\ell m}$ modes \cite{microwave:modes}, as well as
appreciating that some modes help to break particular parameter
degeneracies.  We have only scratched the surface of CMB lensing so far,
and it is expected that future small-scale experiments will be able to extract
more of the cosmological information.
Further constraints can also be derived on the lower-redshift Universe by
cross-correlating CMB lensing with other cosmological tracers of
large-scale structure.
Additionally, small-scale lensing, combined with $E$-mode measurements, can
be used to `delens' CMB $B$-mode data, which will be important for
pushing down into the $r\lesssim0.01$ regime \cite{microwave:KnoxSong,*microwave:Kesdenetal,*microwave:HirataSeljak}.

\section{Complications}
\label{microwave:sec:CMBcomplications}

There are a number of issues that complicate the interpretation of CMB
anisotropy data (and are considered to be {\it signal\/} by many
astrophysicists), some of which we sketch out below.

\subsection{Foregrounds}
\label{microwave:sec:CMBforegrounds}

The microwave sky contains
significant emission from our Galaxy and from extragalactic
sources \cite{microwave:PlanckComponents,*microwave:Planck2015ComponentsII,*microwave:Planck2018Components}.
Fortunately, the frequency dependence of these
various sources is in general substantially different from that of the CMB
anisotropy signals.  The combination of Galactic synchrotron, bremsstrahlung,
and dust emission reaches a minimum at a frequency of roughly $100\,$GHz
(or wavelength of about $3\,$mm).
As one moves to greater angular resolution, the
minimum moves to slightly higher frequencies, but becomes more sensitive
to unresolved (point-like) sources.

At frequencies around $100\,$GHz, and for portions of the sky away from
the Galactic plane, the foregrounds are typically 1 to 10\% of the CMB
anisotropies.  By making observations at multiple frequencies, it is relatively
straightforward to separate the various components and determine the CMB
signal to the few per cent level.  For greater sensitivity, it is necessary to
use the spatial information and statistical properties of the foregrounds
to separate them from the CMB.  Furthermore, at higher $\ell$s it is essential
to carefully model extragalactic foregrounds, particularly the clustering of
infrared-emitting galaxies and scattering due to galaxy clusters,
which dominate the measured power spectrum as
we move into the damping tail.

The foregrounds for CMB polarization follow a
similar pattern to those for temperature, but are
intrinsically brighter relative to CMB anisotropies.  {\it WMAP\/} showed
that the polarized foregrounds dominate at large angular scales, and that they
must be well characterized in order to be
discriminated \cite{microwave:gold11}.
{\it Planck\/} has shown that it is possible to characterize the foreground
polarization signals, with synchrotron dominating at low frequencies and
dust at high frequencies \cite{microwave:PlanckDust}.  On smaller scales there
are no strongly-polarized foregrounds, and hence at high multipoles
it is in principle easier to measure foreground-free modes in polarization than
in temperature.  Although foreground contamination will no doubt
become more complicated as we push down in sensitivity,
making analysis more difficult,
for the time being, foreground contamination is not a
fundamental limit for CMB experiments.

\subsection{Secondary Anisotropies}
\label{microwave:sec:CMBsecondaries}

With increasingly precise measurements of the primary anisotropies, there
is growing theoretical and observational interest in `secondary anisotropies,'
pushing experiments to higher angular resolution and sensitivity.  These
secondary effects arise from the processing of the CMB due to ionization
history and the evolution of structure, including gravitational lensing (which
was already discussed) and
patchy reionization effects \cite{microwave:millea11}.
Additional information can thus
be extracted about the Universe at $z\ll1000$.  This tends to be most
effectively done through correlating CMB maps with other cosmological probes
of structure.  Secondary signals are also typically non-Gaussian, unlike the
primary CMB anisotropies.
\index{Cosmological probes!CMB}%

%\subsection{Sunyaev-Zeldovich Effect}
%\labelsubsection{CMBSZ}
%\IndexEntry{CMBSZ}

\index{Sunyaev-Zeldovich effect}%
\index{SZ, Sunyaev-Zeldovich effect}%
A secondary signal of great current interest is the Sunyaev-Zeldovich
(SZ) effect \cite{microwave:sunzel80}, which is Compton scattering
($\gamma e \rightarrow \gamma' e'$)
of the CMB photons by hot electrons in intergalactic plasma.  This creates
spectral distortions by transferring energy from the electrons to the photons.
It is particularly important for clusters of galaxies, through which
one observes a partially Comptonized spectrum, resulting in a decrement at
radio wavelengths and an increment in the submillimeter.
\index{CMB!Compton distortion}%

The imprint on the CMB sky is of the form $\Delta T/T = y\, f(x)$, with
the $y$ parameter being the integral of Thomson optical depth times
$kT_{\rm e}/m_{\rm e}c^2$ through the cluster, and $f(x)$ describing the
frequency dependence.  This is simply $x\coth(x/2)-4$ for a non-relativistic
gas (the electron temperature in a cluster is typically a few keV),
where the dimensionless frequency $x\equiv h\nu/kT_\gamma$.
As well as this
`thermal' SZ effect, there is also a smaller `kinetic' effect due to the
bulk motion of the cluster gas, giving $\Delta T/T\sim \tau (v/c)$,
with either sign, but having the same spectrum as the primary CMB anisotropies.

A significant advantage in finding galaxy clusters via the SZ effect is
that the signal is largely independent of redshift, so in principle clusters
can be found to arbitrarily large distances.  The SZ effect can be used to
find and study individual clusters, and to obtain estimates of
the Hubble constant.  There is also the potential to constrain other
cosmological parameters, such as the clustering amplitude $\sigma_8$ and
the equation of state of the dark energy, through counts of detected clusters
as a function of redshift.
The promise of the method has been realized through detections of clusters
purely through the SZ effect by SPT \cite{microwave:SPTSZ},
ACT \cite{microwave:ACTSZ2}, and {\it Planck\/} \cite{microwave:PlanckSZ}.
Results from {\it Planck\/} clusters \cite{microwave:PlanckSZCosmology}
suggest a somewhat lower value of $\sigma_8$
than inferred from CMB anisotropies, but there are still systematic
uncertainties that might encompass the difference, and a more recent analysis
of SPT-detected clusters shows better agreement \cite{microwave:DeHaan16}.
Further analysis of
scaling relations among cluster properties should enable more robust
cosmological constraints to be placed in future, so that we can understand
whether this `tension' might be a sign of new physics.

\subsection{Higher-order Statistics}
\label{microwave:sec:CMBstatistics}

Although most of the CMB anisotropy information is contained in the power
spectra, there will also be weak signals present in higher-order statistics.
These can measure any primordial non-Gaussianity in the
perturbations,
as well as non-linear growth of the fluctuations on small scales and other
secondary effects (plus residual foreground contamination of course).
There are an infinite variety of ways in which the CMB could be
non-Gaussian \cite{microwave:bartolo04}; however,
there is a generic form to consider for the initial conditions,
where a quadratic contribution to the curvature perturbations is
parameterized through a dimensionless number $f_{\rm NL}$.  This weakly
non-linear component can be constrained in several ways, the most popular
being through measurements of the bispectrum (or 3-point function).

The constraints depend on the shape of the triangles in harmonic space,
and it has become common to distinguish the `local' or `squeezed'
configuration (in which one side is much smaller than
the other two) from the `equilateral' configuration.  Other configurations
are also relevant for specific theories, such as `orthogonal' non-Gaussianity,
which has positive correlations for $k_1\simeq2k_2\simeq2k_3$,
and negative correlations for the equilateral configuration.
The constraints from the {\it Planck\/} team \cite{microwave:Planck2018NG}
are
$f_{\rm NL}^{\rm local} =1\pm5$,
$f_{\rm NL}^{\rm equil} =-26\pm47$, and
$f_{\rm NL}^{\rm ortho} =-38\pm24$.

These results are consistent with zero, but are at a level that is now
interesting for model predictions.
The amplitude of $f_{\rm NL}$ expected is small, so that a detection of
$f_{\rm NL}\gg1$ would rule out all single-field, slow-roll inflationary
models.  It is still possible to improve upon these {\it Planck\/}
results, and it certainly seems feasible that
a measurement of primordial non-Gaussianity may yet be within reach.
{\it Non}-primordial detections of non-Gaussianity from expected signatures
have already been made.  For example, the bispectrum and trispectrum
contain evidence of gravitational lensing, the ISW effect, and Doppler
boosting.  For now the primordial signal is elusive, but should it be
detected, then detailed measurements of non-Gaussianity will become a
unique probe of inflationary-era physics.
Because of that, much effort continues to be devoted to honing predictions and
measurement techniques, with the expectation that we will need to go beyond
the CMB (\eg,~3D galaxy surveys) to dramatically improve the constraints.

\subsection{Anomalies}
\label{microwave:sec:CMBanomalies}

Several features seen in the {\it Planck\/}
data \cite{microwave:PlanckIandS,microwave:Planck2015IandS,microwave:Planck2018IandS}
confirm those found earlier with {\it WMAP\/} \cite{microwave:WMAPanomalies},
showing mild deviations from a simple description of the sky; these are
often referred to as `anomalies.'  One such feature is the lack of
power in the multipole range
$\ell\simeq20$--30 \cite{microwave:Planck2018Params,microwave:Planck2018Like}.
Other examples involve the apparent breaking of statistical
anisotropy, caused by alignment of the lowest multipoles, as well as a somewhat
excessive cold spot and a power asymmetry between hemispheres.  No such
feature is significant at more than the roughly $3\sigma$ level, and
the importance of `{\it a posteriori\/}'
statistics here has been emphasized by many
authors.  Since these effects are at large angular scales, where cosmic
variance dominates, the results will not increase in significance with more
data, although there is the potential for more sensitive polarization
measurements to provide independent tests.

\section{Constraints on Cosmological Parameters}
\label{microwave:sec:CMBconstraints}

The most striking outcome of the last couple of decades of
experimental results is that the standard cosmological paradigm continues
to be in very good shape.  A large amount of high-precision
data on the CMB power spectrum is adequately fit with fewer than 10
free parameters (and only six need non-trivial values).
The framework is that of FRW
models, which have nearly flat geometry, containing dark matter and dark
energy, and with adiabatic perturbations having close to scale-invariant
initial conditions.

Within this basic picture, the values of the
cosmological parameters can be constrained.  Of course,
more stringent bounds can be placed on models
that cover a restricted parameter space, \eg,~assuming
that ${\rm \Omega}_{\rm tot}=1$ or $r=0$.  More generally, the
constraints depend upon the adopted prior probability distributions,
even if they are implicit,
for example by restricting the parameter freedom or their ranges
(particularly where likelihoods peak near the boundaries), or
by using different choices of other data in combination with the CMB.
As the data become even more precise,
these considerations will be less important,
but for now we caution that restrictions on model space and choice of
non-CMB data sets and priors need to be kept in mind when adopting
specific parameter values and uncertainties.

There are some combinations of parameters that fit the CMB anisotropies
almost equivalently.  For example, there is a nearly exact geometric
degeneracy,
where any combination of ${\rm \Omega}_{\rm m}$ and ${\rm \Omega}_{\rm \Lambda}$
that provides the same angular-diameter distance to last
scattering will give nearly identical $C_\ell$s.  There are also other
less exact degeneracies among the parameters.  Such degeneracies can be broken
when using the CMB results in combination with other cosmological data sets.
Particularly useful are complementary constraints from baryon acoustic
oscillations, galaxy clustering, the abundance of galaxy clusters,
weak gravitational lensing measurements, and Type Ia supernova distances.
For an overview of some of these other cosmological constraints,
see The Cosmological Parameters---Sec.~\ref{hubble} of this
\index{dA, angular-diameter distance@$d_A$, angular-diameter distance}%
{\it Review}.

Within the context of a 6-parameter family of models (which fixes
${\rm \Omega}_{\rm tot}=1$, $dn_{\rm s}/d\ln k=0$, $r=0$, and $w=-1$) the
{\it Planck\/} results for $TT$, together with $TE$, $EE$, and
CMB lensing, yield \cite{microwave:Planck2018Params}:
$\ln(10^{10}A_{\rm s})=3.044\pm0.014$;
$n_{\rm s}=0.965\pm0.004$;
${\rm \Omega}_{\rm b}h^2=0.02237\pm0.00015$;
${\rm \Omega}_{\rm c}h^2=0.1200\pm0.0012$;
$100\theta_\ast=1.04092\pm0.00031$;
and $\tau=0.054\pm0.007$.
Other parameters can be derived from this basic set, including
$h=0.674\pm0.005$, ${\rm \Omega}_{\rm \Lambda}=0.685\pm0.007$ ($=1-{\rm \Omega}_{\rm m}$)
and $\sigma_8=0.811\pm0.006$ (see also Astrophysical Constants and
Parameters---Sec.~2 of the Review of Particle Physics).
Somewhat different (although consistent) values are obtained using other data
combinations, such as including BAO, supernova, $H_0$, or weak-lensing
constraints (see Sec.~\ref{hubble} of the {\it Review of Particle Physics}).  However,
the {\it Planck\/} results quoted above are currently the best available from
CMB data alone.  Results from other CMB experiments
(\eg, SPT-3G \cite{microwave:Balkenhol2023}) are
consistent and becoming competitive.

The standard cosmological model still fits the data well, with the
error bars on the parameters continuing to shrink.
Improved measurement of higher acoustic peaks has dramatically reduced
the uncertainty in the $\theta_\ast$ parameter, which is now detected at
$>3000\sigma$.  The evidence for $n_{\rm s}<1$ is now at the $8\sigma$
level from {\it Planck\/} data alone.
The value of the reionization optical depth has decreased
compared with earlier estimates; it is convincingly detected, but
still not at very high significance.

%Although not in the usual list of six CMB-derived parameters, the Hubble
%constant can be derived from that set.
The inferred value of $H_0$ is
smaller than the most precise values derived from the cosmic distance ladder.
\index{Hubble!constant $H_0$}%
This parameter tension is discussed more fully in other sections of this
{\it Review}.
The CMB anisotropies also provide the most precise estimate of the age of the
Universe, with {\it Planck\/} giving the value $t_0=13.797\pm0.023\,$Gyr.

Constraints can also be placed on parameters beyond the basic six,
particularly when including other astrophysical data sets.
Relaxing the flatness assumption, the constraint on ${\rm \Omega}_{\rm tot}$ is
$1.011\pm0.006$.
Note that for $h$, the CMB data alone provide only a very weak constraint if
spatial flatness is not assumed.
However, with the addition of other data (particularly powerful in this context
being a compilation of BAO measurements; see Sec.~\ref{hubble} of this
{\it Review}),
the constraints on the Hubble constant and curvature improve considerably,
leading to ${\rm \Omega}_{\rm tot}
 = 0.9993\pm0.0019$\cite{microwave:Planck2018Params}.

For ${\rm \Omega}_{\rm b}h^2$ the CMB-derived value is generally consistent
with completely independent constraints from Big Bang nucleosynthesis
(see Sec.~\ref{bigbangnuc} of the {\it Review of Particle Physics}).  Related are
constraints
on additional neutrino-like relativistic degrees of freedom, which lead to
$N_{\rm eff}=2.99\pm0.17$ (including BAO),
\ie,~no evidence for extra neutrino species.

The tightest published limit on the tensor-to-scalar ratio is $r<0.036$
(measured at $k=0.05\,{\rm Mpc}^{-1}$) from BICEP/Keck Array \cite{microwave:BKXIII}.
The detailed limit depends on how other parameters, especially $A_{\rm s}$,
$n_{\rm t}$, and $dn_{\rm s}/d\ln k \ne 0$ are restricted.
The joint constraints on
$n_{\rm s}$ and $r$ allow specific inflationary models to be
tested \cite{microwave:PlanckInfl,*microwave:Planck2015Infl,microwave:Planck2018Infl}.
Looking at the $(n_{\rm s},r)$ plane,
this means that $m^2\phi^2$ (mass-term quadratic) inflation is disfavored
by the data, as well as $\lambda \phi^4$ (self-coupled) inflation.

The addition of the dark-energy equation of state $w$ adds the partial
degeneracy of being able to fit a ridge in $(w,h)$ space, extending to
low values of both parameters.  This degeneracy is broken when the CMB is
used in combination with other data sets, \eg,~adding a compilation of BAO and
supernova data gives $w=-1.028\pm0.031$\cite{microwave:Planck2018Params}.
Constraints can also be placed on more
general dark energy and modified-gravity models \cite{microwave:PlanckDEMG}.
However, when extending the search space,
one needs to be careful not to over-interpret some tensions between
data sets as evidence for new physics.

For the reionization optical depth, a reanalysis of {\it Planck\/} data in 2016
resulted in a reduction in the value of $\tau$, with the tightest result
giving $\tau=0.055\pm0.009$, and the newest analyses giving similar numbers.
This corresponds to $z_{\rm i}=7.8$--8.8 (depending on the functional form of
the reionization history), with an uncertainty of
$\pm0.9$\cite{microwave:Reion16}.  This redshift is only slightly higher that
that suggested from studies of absorption lines in high-$z$ quasar
spectra \cite{microwave:fan06} and
Ly$\,\alpha$-emitting galaxies \cite{microwave:Mason18},
perhaps hinting that the process of reionization was not as complex as
previously suspected.  The important
constraint provided by CMB polarization, in combination with
astrophysical measurements, thus allows us to investigate how
the first stars formed and brought about the end of the cosmic dark ages.

\section{Particle Physics Constraints}
\label{microwave:sec:CMBparticles}

CMB data place limits on parameters that are directly
relevant for particle physics models.  For example,  there is a limit on
the sum of the masses of the neutrinos,
${\rm \Sigma} m_{\nu}<0.12\,$eV (95\%) \cite{microwave:Planck2018Params}
coming from {\it Planck\/} together with BAO measurements (although limits are
weaker when considering both $N_{\rm eff}$ and ${\rm \Sigma} m_{\nu}$ as free
parameters).  This assumes the usual number density of fermions,
which decoupled when they were relativistic.  The limit is tantalizingly only
a factor of a few higher than the minimum value coming from neutrino mixing
experiments (see Neutrino Mixings---Secs.~\ref{numix} and
\ref{nucosm}).  As well as being an
indirect probe of the neutrino background, {\it Planck\/} data
also require that the neutrino fluid has perturbations, \ie,~that it
possesses a sound speed $c_{\rm s}^2\simeq 1/3$,
as expected \cite{microwave:Planck2015Params}.

The current suite of data suggests that $n_{\rm s}<1$,
with a best-fitting value about 0.035 below unity.  This is already
quite constraining for inflationary models, particularly along with $r$
limits.  There is no current evidence for running of the
spectral index, with $dn_{\rm s}/d\ln k = -0.004\pm0.007$
from {\it Planck\/} alone \cite{microwave:Planck2018Params}
(with a similar value when BAO data are included),
although this is less of a constraint on models.
Similarly, primordial non-Gaussianity is being probed to interesting levels,
although tests of simple inflationary models will only come with significant
reductions in uncertainty.

The large-angle anomalies, such as the hemispheric modulation of power and
the dip in power at $\ell\simeq20$--30,
have the potential to be hints of new physics.
Such effects might be expected in a Universe that has
a large-scale power cut-off, or anisotropy in the initial power spectrum, or is
topologically non-trivial.  However, cosmic variance and {\it a posteriori\/}
statistics limit the significance of these anomalies, absent the existence of
a model that naturally yields some of these features (and ideally also
predicting other phenomena that can be tested).

Constraints on `cosmic birefringence' (\ie,~rotation of the plane of CMB
polarization that generates non-zero $TB$ and $EB$ power) can be used to
place limits on theories involving parity violation, Lorentz violation,
or
\index{Axion-photon mixing}%
axion-photon mixing \cite{microwave:PlanckParity,*microwave:Komatsu2022}.

It is possible to place limits on additional areas of
physics \cite{microwave:KamKos}, for example
\index{Annihilating dark matter}%
\index{Dark matter!annihilating}%
annihilating dark matter \cite{microwave:Planck2015Params,microwave:Planck2015Params},
\index{Primordial magnetic fields}%
\index{Magnetic fields, primordial}%
primordial magnetic fields \cite{microwave:PlanckMagnetic},
and
\index{Time variation of the fine-structure constant}%
\index{Fine-structure constant, time variation}%
time variation of the fine-structure constant \cite{microwave:PlanckConstants},
as well as the
\index{Neutrino(s)!chemical potential}%
\index{Chemical potential, neutrino}%
neutrino chemical potential, a contribution
of warm dark matter, topological defects,
or
\index{Physics beyond general relativity}%
\index{Beyond the general relativity physics}%
physics beyond general relativity.  Further particle physics constraints
will follow as the
smaller-scale and polarization measurements continue to improve.

The CMB anisotropy measurements precisely pin down physics at the time of
last-scattering, and so any change of physics can be constrained if it
affects the relevant energies or timescales.  Future, higher sensitivity
measurements of the CMB frequency spectrum will push the constraints back to
cover energy injection at much earlier times ($\sim1\,$year).  Comparison of
CMB and BBN observables extend these constraints to timescales of order
seconds, and energies in the MeV range.  And to the extent that inflation
provides an effective description of the generation of perturbations, the
inflationary observables may constrain physics at GUT-type energy scales.

More generally, careful
measurement of the CMB power spectra and non-Gaussianity can in
principle put constraints on physics at the highest energies, including ideas
of quantum gravity, string theory, extra dimensions, colliding branes,
{\it etc}.  At the moment
any calculation of predictions appears to be far from definitive.  However,
there is a great deal of activity on implications of fundamental theories for
the early Universe, and hence a chance that there might be
observational implications for specific scenarios.

\section{Fundamental Lessons}
\label{microwave:sec:CMBlessons}

More important than the precise values of parameters is what we have
learned about the general features that describe our observable Universe.
Beyond the basic hot Big Bang picture, the CMB has taught us that:

\begin{itemize}
\item
the (observable) Universe is very close to isotropic;
\item
the Universe recombined at $z\sim1000$ and started to become ionized again at
$z\sim10$;
\item
the geometry of the Universe is close to flat;
\item
both dark matter and dark energy are required;
\item
gravitational instability is sufficient to grow all of the observed large
structures in the Universe;
\item
topological defects were not important for structure formation;
\item
there were `synchronized' super-Hubble modes generated in the early
Universe;
\item
the initial perturbations were predominantly adiabatic in nature;
\item
the primordial perturbation spectrum has a slightly red tilt;
\item
the perturbations had close to Gaussian (\ie,~maximally random)
initial conditions.
\end{itemize}

These features form the basis of the
\index{Standard model, cosmological}%
\index{Cosmological!standard model}%
cosmological standard model,
${\rm \Lambda}$CDM,
for which it is tempting to make an analogy with the Standard Model of
particle physics (see earlier Sections of the {\it Review of Particle Physics}).
In comparison, the cosmological model is much further from any underlying
`fundamental theory,' which might ultimately provide the
values of the parameters from first principles.  Nevertheless, any
genuinely complete `theory of everything' must include an explanation for
the values of these cosmological parameters in addition to the parameters of
the Standard Model of particle physics.

\section{Future Directions}
\label{microwave:sec:CMBfuture}

Given the significant progress in measuring the CMB sky,
which has been instrumental in tying down the
cosmological model, what can we anticipate for the future?
There will
be a steady improvement in the precision and confidence with which we
can determine the appropriate cosmological parameters.
Ground-based experiments operating at smaller
angular scales will continue to place tighter constraints on the damping tail,
lensing, and cross-correlations.  New polarization experiments at small
scales will probe further into the damping tail, without the
limitation of extragalactic foregrounds.  And polarization experiments at
large angular scales will push down the limits on primordial $B$ modes.

{\it Planck}, the third generation CMB satellite mission, was launched in
May 2009, and produced a large number of papers,
including a set of cosmological studies based on the first two full
surveys of the sky (accompanied by a public release of data products)
in 2013, a further series coming from analysis of the full mission
data release in 2015 (eight surveys for the Low Frequency Instrument and
five surveys for the High Frequency Instrument), and a third series derived
from a final analysis of the 2018 data release, including
full constraints from polarization data.  {\it Planck\/} data currently
dominate constraints on models, but that situation will change soon.

A set of cosmological parameters is now known to percent-level accuracy,
and that may seem sufficient for many people.
However, we should certainly demand more of measurements that describe
{\it the entire observable Universe\/!}  Hence a lot of activity in the coming
years will continue to focus on determining those parameters with
increasing precision.  This necessarily includes testing for consistency
among different predictions of the cosmological Standard Model, and searching
for signals that might require additional physics.

A second area of focus will be the smaller-scale anisotropies and
`secondary effects.'  There is a great deal of information about structure
formation at $z\ll1000$ encoded in the CMB sky.  This may involve
higher-order statistics and cross-correlations with other large-scale
structure tracers, as well as spectral signatures, with many
experiments targeting the galaxy cluster SZ effect, for example.
The current status of
CMB lensing is similar (in terms of total signal-to-noise) to the quality of
the first CMB anisotropy measurements by {\it COBE}, and thus we can expect
that experimental probes of lensing will improve dramatically in the coming
years.  All of these investigations
can provide constraints on the dark-energy equation of state, for example,
which is a major area of focus for several future cosmological surveys at
optical wavelengths.  CMB lensing also promises to yield a measurement of
the sum of the neutrino masses.

A third direction is increasingly sensitive searches for specific signatures of
physics at the highest energies.  The most promising of these may be the
primordial gravitational wave signals in $C_\ell^{BB}$, which could
be a probe of the $\sim10^{16}\,$GeV energy range.
There are several experiments underway or being developed that are designed to
search for the polarization $B$ modes, with the most ambitious being CMB-S4 on
the ground and {\it LiteBIRD\/} in space.
Additionally, non-Gaussianity holds the promise of
constraining models beyond single-field slow-roll inflation.

Anisotropies in the CMB have proven to be the premier probe of cosmology
and the early Universe.  Theoretically the CMB involves well-understood
physics in the linear regime, and is under very good calculational
control.  A substantial and improving set of observational data now exists.
Systematics appear to be under control and are not currently a limiting factor.
And so for the next several years we can expect an increasing amount of
cosmological information to be gleaned from CMB anisotropies, with the
prospect also of some genuine surprises.

\bibliographystyle{unsrt}
\bibliography{microwave.bib}

\begin{thebibliography}{100}

\bibitem{microwave:Penzias65}
A.~A. {Penzias} and R.~W. {Wilson}.
\newblock {A Measurement of Excess Antenna Temperature at 4080 Mc/s.}
\newblock {\em Astrophys. J.}, 142:419--421, July 1965.

\bibitem{microwave:Dicke}
R.~H. {Dicke}, P.~J.~E. {Peebles}, P.~G. {Roll}, and D.~T. {Wilkinson}.
\newblock {Cosmic Black-Body Radiation.}
\newblock {\em Astrophys. J.}, 142:414--419, July 1965.

\bibitem{microwave:WSS}
M.~{White}, D.~{Scott}, and J.~{Silk}.
\newblock {Anisotropies in the Cosmic Microwave Background}.
\newblock {\em Ann. Rev. Astron. Astrophys.}, 32:319--370, 1994.

\bibitem{microwave:HD}
W.~{Hu} and S.~{Dodelson}.
\newblock {Cosmic Microwave Background Anisotropies}.
\newblock {\em Ann. Rev. Astron. Astrophys.}, 40:171--216, 2002.

\bibitem{microwave:CP}
A.~{Challinor} and H.~{Peiris}.
\newblock {Lecture notes on the physics of cosmic microwave background
  anisotropies}.
\newblock In M.~{Novello} and S.~{Perez}, editors, {\em American Institute of
  Physics Conference Series}, volume 1132, pages 86--140, May 2009.

\bibitem{microwave:smoot92}
G.~F. {Smoot}, C.~L. {Bennett}, A.~{Kogut}, E.~L. {Wright}, J.~{Aymon}, N.~W.
  {Boggess}, E.~S. {Cheng}, G.~{de Amici}, et~al.
\newblock {Structure in the COBE differential microwave radiometer first-year
  maps}.
\newblock {\em Astrophys. J. Lett.}, 396:L1--L5, September 1992.

\bibitem{microwave:bennett03}
C.~L. {Bennett}, M.~{Halpern}, G.~{Hinshaw}, N.~{Jarosik}, A.~{Kogut},
  M.~{Limon}, S.~S. {Meyer}, L.~{Page}, et~al.
\newblock {First-Year Wilkinson Microwave Anisotropy Probe (WMAP) Observations:
  Preliminary Maps and Basic Results}.
\newblock {\em Astrophys. J. Supp.}, 148:1--27, September 2003.

\bibitem{microwave:hinshaw12}
G.~{Hinshaw}, D.~{Larson}, E.~{Komatsu}, D.~N. {Spergel}, C.~L. {Bennett},
  J.~{Dunkley}, M.~R. {Nolta}, M.~{Halpern}, et~al.
\newblock {Nine-year Wilkinson Microwave Anisotropy Probe (WMAP) Observations:
  Cosmological Parameter Results}.
\newblock {\em Astrophys. J. Supp.}, 208:19, October 2013.

\bibitem{microwave:PlanckParams}
{Planck Collab.\ 2013 Results XVI}.
\newblock {Planck 2013 results. XVI. Cosmological parameters}.
\newblock {\em Astron. Astrophys.}, 571:A16, November 2014.

\bibitem{microwave:tauber}
J.~A. {Tauber}, N.~{Mandolesi}, J.-L. {Puget}, T.~{Banos}, M.~{Bersanelli},
  F.~R. {Bouchet}, R.~C. {Butler}, J.~{Charra}, G.~{Crone}, J.~{Dodsworth}, and
  et~al.
\newblock {Planck pre-launch status: The Planck mission}.
\newblock {\em Astron. Astrophys.}, 520:A1, September 2010.

\bibitem{microwave:Planck2013I}
{Planck Collab.\ 2013 Results I}.
\newblock {Planck 2013 results. I. Overview of products and scientific
  results}.
\newblock {\em Astron. Astrophys.}, 571:A1, November 2014.

\bibitem{microwave:Planck2015I}
{Planck Collab.\ 2015 Results I}.
\newblock {Planck 2015 results. I. Overview of products and scientific
  results}.
\newblock {\em Astron. Astrophys.}, 594:A1, September 2016.

\bibitem{microwave:Planck2015Params}
{Planck Collab.\ 2015 Results XIII}.
\newblock {Planck 2015 results. XIII. Cosmological parameters}.
\newblock {\em Astron. Astrophys.}, 594:A13, September 2016.

\bibitem{microwave:Planck2018I}
{Planck Collab.\ 2018 Results I}.
\newblock {Planck 2018 results. I. Overview and the cosmological legacy of
  Planck}.
\newblock {\em Astron.\ Astrophys.}, 641:A1, September 2020.

\bibitem{microwave:Planck2018Params}
{Planck Collab.\ 2018 Results VI}.
\newblock {Planck 2018 results. VI. Cosmological parameters}.
\newblock {\em Astron. Astrophys.}, 641:A6, September 2020.

\bibitem{microwave:Swetz11}
D.~S. {Swetz}, P.~A.~R. {Ade}, M.~{Amiri}, J.~W. {Appel}, E.~S. {Battistelli},
  B.~{Burger}, J.~{Chervenak}, M.~J. {Devlin}, et~al.
\newblock {Overview of the Atacama Cosmology Telescope: Receiver,
  Instrumentation, and Telescope Systems}.
\newblock {\em Astrophys. J. Supp.}, 194:41, June 2011.

\bibitem{microwave:Carlstrom11}
J.~E. {Carlstrom}, P.~A.~R. {Ade}, K.~A. {Aird}, B.~A. {Benson}, L.~E. {Bleem},
  S.~{Busetti}, C.~L. {Chang}, E.~{Chauvin}, et~al.
\newblock {The 10 Meter South Pole Telescope}.
\newblock {\em Proc. Astron. Soc. Pacific}, 123:568, May 2011.

\bibitem{microwave:Klimenko2020}
V.~V. {Klimenko}, A.~V. {Ivanchik}, P.~{Petitjean}, P.~{Noterdaeme}, and
  R.~{Srianand}.
\newblock {Estimation of the Cosmic Microwave Background Temperature from
  Atomic C I and Molecular CO Lines in the Interstellar Medium of Early
  Galaxies}.
\newblock {\em Astronomy Letters}, 46(11):715--725, November 2020.

\bibitem{microwave:Riechers2022}
Dominik~A. {Riechers}, Axel {Weiss}, Fabian {Walter}, Christopher~L. {Carilli},
  Pierre {Cox}, Roberto {Decarli}, and Roberto {Neri}.
\newblock {Microwave background temperature at a redshift of 6.34 from H$_{2}$O
  absorption}.
\newblock {\em Nature}, 602(7895):58--62, February 2022.

\bibitem{microwave:ARCADE}
D.~J. {Fixsen}, A.~{Kogut}, S.~{Levin}, M.~{Limon}, P.~{Lubin}, P.~{Mirel},
  M.~{Seiffert}, J.~{Singal}, et~al.
\newblock {ARCADE 2 Measurement of the Absolute Sky Brightness at 3-90 GHz}.
\newblock {\em Astrophys. J.}, 734:5, June 2011.

\bibitem{microwave:Singal18}
J.~{Singal}, J.~{Haider}, M.~{Ajello}, D.~R. {Ballantyne}, E.~{Bunn},
  J.~{Condon}, J.~{Dowell}, D.~{Fixsen}, N.~{Fornengo}, B.~{Harms},
  G.~{Holder}, E.~{Jones}, K.~{Kellermann}, A.~{Kogut}, T.~{Linden},
  R.~{Monsalve}, P.~{Mertsch}, E.~{Murphy}, E.~{Orlando}, M.~{Regis},
  D.~{Scott}, T.~{Vernstrom}, and L.~{Xu}.
\newblock {The Radio Synchrotron Background: Conference Summary and Report}.
\newblock {\em Proc.\ Astron.\ Soc.\ Pacific}, 130(985):036001, Mar 2018.

\bibitem{microwave:PIXIE}
A.~{Kogut}, D.~J. {Fixsen}, D.~T. {Chuss}, J.~{Dotson}, E.~{Dwek},
  M.~{Halpern}, G.~F. {Hinshaw}, S.~M. {Meyer}, et~al.
\newblock {The Primordial Inflation Explorer (PIXIE): a nulling polarimeter for
  cosmic microwave background observations}.
\newblock {\em J. Cosmology Astropart. Phys.}, 2011(7):025, July 2011.

\bibitem{microwave:PRISM}
Philippe {Andr{\'e}}, Carlo {Baccigalupi}, Anthony {Banday}, Domingos
  {Barbosa}, Belen {Barreiro}, James {Bartlett}, Nicola {Bartolo}, Elia
  {Battistelli}, et~al.
\newblock {PRISM (Polarized Radiation Imaging and Spectroscopy Mission): an
  extended white paper}.
\newblock {\em J. Cosmology Astropart. Phys.}, 2014(2):006, Feb 2014.

\bibitem{microwave:Delabrouille21}
J.~{Delabrouille}, M.~H. {Abitbol}, N.~{Aghanim}, Y.~{Ali-Haimoud},
  D.~{Alonso}, M.~{Alvarez}, A.~J. {Banday}, J.~G. {Bartlett}, et~al.
\newblock {Microwave Spectro-Polarimetry of Matter and Radiation across Space
  and Time}.
\newblock {\em Experimental Astronomy}, 51(3):1471--1514, June 2021.

\bibitem{microwave:Desjacques15}
V.~{Desjacques}, J.~{Chluba}, J.~{Silk}, F.~{de Bernardis}, and O.~{Dor{\'e}}.
\newblock {Detecting the cosmological recombination signal from space}.
\newblock {\em Mon. Not. R. Astron. Soc.}, 451:4460--4470, August 2015.

\bibitem{microwave:fixsen09}
D.~J. {Fixsen}.
\newblock {The Temperature of the Cosmic Microwave Background}.
\newblock {\em Astrophys. J.}, 707:916--920, December 2009.

\bibitem{microwave:mather99}
J.~C. {Mather}, D.~J. {Fixsen}, R.~A. {Shafer}, C.~{Mosier}, and D.~T.
  {Wilkinson}.
\newblock {Calibrator Design for the COBE Far-Infrared Absolute
  Spectrophotometer (FIRAS)}.
\newblock {\em Astrophys. J.}, 512:511--520, February 1999.

\bibitem{microwave:Hoffman15}
Y.~{Hoffman}, H.~M. {Courtois}, and R.~B. {Tully}.
\newblock {Cosmic bulk flow and the local motion from Cosmicflows-2}.
\newblock {\em Mon. Not. R. Astron. Soc.}, 449:4494--4505, June 2015.

\bibitem{microwave:fixsen94}
D.~J. {Fixsen}, E.~S. {Cheng}, D.~A. {Cottingham}, R.~E. {Eplee}, Jr., R.~B.
  {Isaacman}, J.~C. {Mather}, S.~S. {Meyer}, P.~D. {Noerdlinger}, et~al.
\newblock {Cosmic microwave background dipole spectrum measured by the COBE
  FIRAS instrument}.
\newblock {\em Astrophys. J.}, 420:445--449, January 1994.

\bibitem{microwave:Planck2013XXVII}
{Planck Collab.\ 2013 Results XXVII}.
\newblock {Planck 2013 results. XXVII. Doppler boosting of the CMB: Eppur si
  muove}.
\newblock {\em Astron. Astrophys.}, 571:A27, November 2014.

\bibitem{microwave:SSS}
S.~{Seager}, D.~D. {Sasselov}, and D.~{Scott}.
\newblock {How Exactly Did the Universe Become Neutral?}
\newblock {\em Astrophys. J. Supp.}, 128:407--430, June 2000.

\bibitem{Knox:1995dq}
LLoyd Knox.
\newblock {Determination of inflationary observables by cosmic microwave
  background anisotropy experiments}.
\newblock {\em Phys. Rev.}, D52:4307--4318, 1995.

\bibitem{microwave:bartolo04}
N.~{Bartolo}, E.~{Komatsu}, S.~{Matarrese}, and A.~{Riotto}.
\newblock {Non-Gaussianity from inflation: theory and observations}.
\newblock {\em Phys. Rep.}, 402:103--266, November 2004.

\bibitem{microwave:Planck2013XXIV}
{Planck Collab.\ 2013 Results XXIV}.
\newblock {Planck 2013 results. XXIV. Constraints on primordial
  non-Gaussianity}.
\newblock {\em Astron. Astrophys.}, 571:A24, November 2014.

\bibitem{microwave:WMAPanomalies}
C.~L. {Bennett}, R.~S. {Hill}, G.~{Hinshaw}, D.~{Larson}, K.~M. {Smith},
  J.~{Dunkley}, B.~{Gold}, M.~{Halpern}, et~al.
\newblock {Seven-year Wilkinson Microwave Anisotropy Probe (WMAP) Observations:
  Are There Cosmic Microwave Background Anomalies?}
\newblock {\em Astrophys. J. Supp.}, 192:17, February 2011.

\bibitem{microwave:PlanckIandS}
{Planck Collab.\ 2013 Results XXIII}.
\newblock {Planck 2013 results. XXIII. Isotropy and statistics of the CMB}.
\newblock {\em Astron. Astrophys.}, 571:A23, November 2014.

\bibitem{microwave:Planck2018Infl}
{Planck Collab.\ 2018 Results X}.
\newblock {Planck 2018 results. X. Constraints on inflation}.
\newblock {\em Astron. Astrophys.}, 641:A10, September 2020.

\bibitem{microwave:lidlyth}
A.~R. {Liddle} and D.~H. {Lyth}.
\newblock {\em {Cosmological Inflation and Large-Scale Structure}}.
\newblock {Cambridge University Press}, {Cambridge}, June 2000.

\bibitem{microwave:SelZal}
U.~{Seljak} and M.~{Zaldarriaga}.
\newblock {A Line-of-Sight Integration Approach to Cosmic Microwave Background
  Anisotropies}.
\newblock {\em Astrophys. J.}, 469:437, October 1996.

\bibitem{microwave:LewChaLas}
A.~{Lewis}, A.~{Challinor}, and A.~{Lasenby}.
\newblock {Efficient Computation of Cosmic Microwave Background Anisotropies in
  Closed Friedmann-Robertson-Walker Models}.
\newblock {\em Astrophys. J.}, 538:473--476, August 2000.

\bibitem{microwave:class}
D.~{Blas}, J.~{Lesgourgues}, and T.~{Tram}.
\newblock {The Cosmic Linear Anisotropy Solving System (CLASS). Part II:
  Approximation schemes}.
\newblock {\em J. Cosmology Astropart. Phys.}, 7:034, July 2011.

\bibitem{Seljak:2003th}
Uros Seljak, Naoshi Sugiyama, Martin~J. White, and Matias Zaldarriaga.
\newblock {A Comparison of cosmological Boltzmann codes: Are we ready for high
  precision cosmology?}
\newblock {\em Phys. Rev.}, D68:083507, 2003.

\bibitem{microwave:SacWol}
R.~K. {Sachs} and A.~M. {Wolfe}.
\newblock {Perturbations of a Cosmological Model and Angular Variations of the
  Microwave Background}.
\newblock {\em Astrophys. J.}, 147:73, January 1967.

\bibitem{microwave:iswcorr}
R.~G. {Crittenden} and N.~{Turok}.
\newblock {Looking for a Cosmological Constant with the Rees-Sciama Effect}.
\newblock {\em Phys. Rev. Lett.}, 76:575--578, January 1996.

\bibitem{microwave:Planck2015ISW}
{Planck Collab.\ 2015 Results XXI}.
\newblock {Planck 2015 results. XXI. The integrated Sachs-Wolfe effect}.
\newblock {\em Astron. Astrophys.}, 594:A21, September 2016.

\bibitem{microwave:HETW}
W.~Hu, D.~J. Eisenstein, M.~Tegmark, and M.~White.
\newblock {Observationally determining the properties of dark matter}.
\newblock {\em Phys. Rev.}, D59(2):023512, December 1998.

\bibitem{microwave:HuSugiyamaSilk}
W.~{Hu}, N.~{Sugiyama}, and J.~{Silk}.
\newblock {The physics of microwave background anisotropies}.
\newblock {\em Nature}, 386:37--43, March 1997.

\bibitem{microwave:PeeYu}
P.~J.~E. Peebles and J.~T. Yu.
\newblock {Primeval Adiabatic Perturbation in an Expanding Universe}.
\newblock {\em Astrophys. J.}, 162:815, December 1970.

\bibitem{microwave:SunZel70}
{Sunyaev, R.~A.} and {Zeldovich, \relax{Ya}.~B.}
\newblock {Small-Scale Fluctuations of Relic Radiation}.
\newblock {\em Astron. Astrophys. Supp.}, 7:3--19, April 1970.

\bibitem{microwave:scott95}
D.~Scott, J.~Silk, and M.~White.
\newblock {From Microwave Anisotropies to Cosmology}.
\newblock {\em Science}, 268:829--835, May 1995.

\bibitem{microwave:eisenstein05}
D.~J. Eisenstein.
\newblock {Dark energy and cosmic sound [review article]}.
\newblock {\em New Astron. Rev.}, 49:360--365, November 2005.

\bibitem{microwave:silk}
J.~Silk.
\newblock {Cosmic Black-Body Radiation and Galaxy Formation}.
\newblock {\em Astrophys. J.}, 151:459, February 1968.

\bibitem{Zaldarriaga:1998ar}
Matias Zaldarriaga and Uros Seljak.
\newblock {Gravitational lensing effect on cosmic microwave background
  polarization}.
\newblock {\em Phys. Rev.}, D58:023003, 1998.

\bibitem{microwave:PlanckLens}
{Planck Collab.\ 2013 Result XVII}.
\newblock {Planck 2013 results. XVII. Gravitational lensing by large-scale
  structure}.
\newblock {\em Astron. Astrophys.}, 571:A17, November 2014.

\bibitem{microwave:kaplighat03}
M.~{Kaplinghat}, L.~{Knox}, and Y.-S. {Song}.
\newblock {Determining Neutrino Mass from the Cosmic Microwave Background
  Alone}.
\newblock {\em Phys. Rev. Lett.}, 91(24):241301, December 2003.

\bibitem{microwave:PlanckLike}
{Planck Collab.\ 2013 Results XV}.
\newblock {Planck 2013 results. XV. CMB power spectra and likelihood}.
\newblock {\em Astron. Astrophys.}, 571:A15, November 2014.

\bibitem{microwave:Planck2018Like}
{Planck Collab.\ 2018 Results V}.
\newblock {Planck 2018 results. V. CMB power spectra and likelihoods}.
\newblock {\em Astron. Astrophys.}, 641:A5, September 2020.

\bibitem{microwave:das13}
S.~{Das}, T.~{Louis}, M.~R. {Nolta}, G.~E. {Addison}, E.~S. {Battistelli},
  J.~R. {Bond}, E.~{Calabrese}, D.~{Crichton}, et~al.
\newblock {The Atacama Cosmology Telescope: temperature and gravitational
  lensing power spectrum measurements from three seasons of data}.
\newblock {\em J. Cosmology Astropart. Phys.}, 4:014, April 2014.

\bibitem{microwave:story13}
K.~T. {Story}, C.~L. {Reichardt}, Z.~{Hou}, R.~{Keisler}, K.~A. {Aird}, B.~A.
  {Benson}, L.~E. {Bleem}, J.~E. {Carlstrom}, et~al.
\newblock {A Measurement of the Cosmic Microwave Background Damping Tail from
  the 2500-Square-Degree SPT-SZ Survey}.
\newblock {\em Astrophys. J.}, 779:86, December 2013.

\bibitem{microwave:hu97}
W.~{Hu} and M.~{White}.
\newblock {A CMB polarization primer}.
\newblock {\em New Astron.}, 2:323--344, October 1997.

\bibitem{Hu:1997hp}
Wayne Hu and Martin~J. White.
\newblock {CMB anisotropies: Total angular momentum method}.
\newblock {\em Phys. Rev.}, D56:596--615, 1997.

\bibitem{Zaldarriaga:1996xe}
Matias Zaldarriaga and Uros Seljak.
\newblock {An all sky analysis of polarization in the microwave background}.
\newblock {\em Phys. Rev.}, D55:1830--1840, 1997.

\bibitem{microwave:KKSpol}
M.~{Kamionkowski}, A.~{Kosowsky}, and A.~{Stebbins}.
\newblock {Statistics of cosmic microwave background polarization}.
\newblock {\em Phys. Rev.}, D55:7368--7388, June 1997.

\bibitem{microwave:dasipol}
J.~M. {Kovac}, E.~M. {Leitch}, C.~{Pryke}, J.~E. {Carlstrom}, N.~W.
  {Halverson}, and W.~L. {Holzapfel}.
\newblock {Detection of polarization in the cosmic microwave background using
  DASI}.
\newblock {\em Nature}, 420:772--787, December 2002.

\bibitem{microwave:larson11}
D.~{Larson}, J.~{Dunkley}, G.~{Hinshaw}, E.~{Komatsu}, M.~R. {Nolta}, C.~L.
  {Bennett}, B.~{Gold}, M.~{Halpern}, et~al.
\newblock {Seven-year Wilkinson Microwave Anisotropy Probe (WMAP) Observations:
  Power Spectra and WMAP-derived Parameters}.
\newblock {\em Astrophys. J. Supp.}, 192:16, February 2011.

\bibitem{microwave:BKV}
{Keck Array and BICEP2 Collabs.\ V}.
\newblock {BICEP2/Keck Array V: Measurements of B-mode Polarization at Degree
  Angular Scales and 150 GHz by the Keck Array}.
\newblock {\em Astrophys. J.}, 811:126, October 2015.

\bibitem{microwave:Choi20}
S.~K. {Choi}, M.~{Hasselfield}, S.-P.~P. {Ho}, B.~{Koopman}, M.~{Lungu}, M.~H.
  {Abitbol}, G.~E. {Addison}, P.~A.~R. {Ade}, et~al.
\newblock {The Atacama Cosmology Telescope: a measurement of the Cosmic
  Microwave Background power spectra at 98 and 150 GHz}.
\newblock {\em J. Cosmology Astropart. Phys.}, 2020(12):045, December 2020.

\bibitem{microwave:SPTPol}
A.~T. {Crites}, J.~W. {Henning}, P.~A.~R. {Ade}, K.~A. {Aird}, J.~E.
  {Austermann}, J.~A. {Beall}, A.~N. {Bender}, B.~A. {Benson}, et~al.
\newblock {Measurements of E-Mode Polarization and Temperature-E-Mode
  Correlation in the Cosmic Microwave Background from 100 Square Degrees of
  SPTpol Data}.
\newblock {\em Astrophys. J.}, 805:36, May 2015.

\bibitem{microwave:Balkenhol2023}
L.~{Balkenhol}, D.~{Dutcher}, A.~{Spurio Mancini}, A.~{Doussot}, K.~{Benabed},
  S.~{Galli}, P.~A.~R. {Ade}, A.~J. {Anderson}, et~al.
\newblock {Measurement of the CMB temperature power spectrum and constraints on
  cosmology from the SPT-3G 2018 TT, TE , and EE dataset}.
\newblock {\em Phys. Rev.}, D108(2):023510, July 2023.

\bibitem{microwave:Planck2018IandS}
{Planck Collab.\ 2018 Results VII}.
\newblock {Planck 2018 results. VII. Isotropy and statistics of the CMB}.
\newblock {\em Astron. Astrophys.}, 641:A7, September 2020.

\bibitem{microwave:Tristram21}
M.~{Tristram}, A.~J. {Banday}, K.~M. {G{\'o}rski}, R.~{Keskitalo}, C.~R.
  {Lawrence}, K.~J. {Andersen}, R.~B. {Barreiro}, J.~{Borrill}, et~al.
\newblock {Planck constraints on the tensor-to-scalar ratio}.
\newblock {\em Astron. Astrophys.}, 647:A128, March 2021.

\bibitem{microwave:Hanson13}
D.~{Hanson}, S.~{Hoover}, A.~{Crites}, P.~A.~R. {Ade}, K.~A. {Aird}, J.~E.
  {Austermann}, J.~A. {Beall}, A.~N. {Bender}, et~al.
\newblock {Detection of B-Mode Polarization in the Cosmic Microwave Background
  with Data from the South Pole Telescope}.
\newblock {\em Phys. Rev. Lett.}, 111(14):141301, October 2013.

\bibitem{microwave:BICEP2}
{BICEP2 Collab.}
\newblock {Detection of B-Mode Polarization at Degree Angular Scales by
  BICEP2}.
\newblock {\em Phys. Rev. Lett.}, 112(24):241101, June 2014.

\bibitem{microwave:BKP}
{BICEP2/Keck and Planck Collabs.}
\newblock {Joint Analysis of BICEP2/Keck Array and Planck Data}.
\newblock {\em Phys. Rev. Lett.}, 114(10):101301, March 2015.

\bibitem{microwave:BK18}
{BICEP2 and Keck Array Collab.}
\newblock {Constraints on Primordial Gravitational Waves Using Planck, WMAP,
  and New BICEP2/Keck Observations through the 2015 Season}.
\newblock {\em Phys. Rev. Lett.}, 121(22):221301, Nov 2018.

\bibitem{microwave:BKXIII}
{BICEP/Keck Collab.}
\newblock {BICEP / Keck XIII: Improved Constraints on Primordial Gravitational
  Waves using Planck, WMAP, and BICEP/Keck Observations through the 2018
  Observing Season}.
\newblock {\em Phys. Rev. Lett.}, 127:151301, October 2021.

\bibitem{microwave:POLARBEAR}
{POLARBEAR Collab.}
\newblock {A Measurement of the Cosmic Microwave Background B-mode Polarization
  Power Spectrum at Subdegree Scales from Two Years of polarbear Data}.
\newblock {\em Astrophys. J.}, 848:121, October 2017.

\bibitem{microwave:Sayre2020}
J.~T. {Sayre}, C.~L. {Reichardt}, J.~W. {Henning}, P.~A.~R. {Ade}, A.~J.
  {Anderson}, J.~E. {Austermann}, et~al.
\newblock {Measurements of B -mode polarization of the cosmic microwave
  background from 500 square degrees of SPTpol data}.
\newblock {\em \prd}, 101(12):122003, June 2020.

\bibitem{microwave:Planck2015Lens}
{Planck Collab.\ 2015 Results XV}.
\newblock {Planck 2015 results. XV. Gravitational lensing}.
\newblock {\em Astron. Astrophys.}, 594:A15, September 2016.

\bibitem{microwave:vanEngelen15}
A.~{van Engelen}, B.~D. {Sherwin}, N.~{Sehgal}, G.~E. {Addison}, R.~{Allison},
  N.~{Battaglia}, F.~{de Bernardis}, J.~R. {Bond}, et~al.
\newblock {The Atacama Cosmology Telescope: Lensing of CMB Temperature and
  Polarization Derived from Cosmic Infrared Background Cross-correlation}.
\newblock {\em Astrophys. J.}, 808:7, July 2015.

\bibitem{microwave:SimonsO}
Peter {Ade}, James {Aguirre}, Zeeshan {Ahmed}, Simone {Aiola}, Aamir {Ali},
  David {Alonso}, Marcelo~A. {Alvarez}, Kam {Arnold}, Peter {Ashton}, Jason
  {Austermann}, Humna {Awan}, Carlo {Baccigalupi}, et~al.
\newblock {The Simons Observatory: science goals and forecasts}.
\newblock {\em J. Cosmology Astropart. Phys.}, 2019(2):056, Feb 2019.

\bibitem{microwave:CMBS4}
K.~N. {Abazajian}, P.~{Adshead}, Z.~{Ahmed}, S.~W. {Allen}, D.~{Alonso}, K.~S.
  {Arnold}, C.~{Baccigalupi}, J.~G. {Bartlett}, et~al.
\newblock {CMB-S4 Science Book, First Edition}.
\newblock {\em ArXiv e-prints}, October 2016.

\bibitem{microwave:LiteBIRD23}
{LiteBIRD Collaboration}.
\newblock {Probing cosmic inflation with the LiteBIRD cosmic microwave
  background polarization survey}.
\newblock {\em Progress of Theoretical and Experimental Physics},
  2023(4):042F01, April 2023.

\bibitem{microwave:Planck2018Lens}
{Planck Collab.\ 2018 Results VIII}.
\newblock {Planck 2018 results. VIII. Gravitational lensing}.
\newblock {\em Astron. Astrophys.}, 641:A8, September 2020.

\bibitem{microwave:Qu2023}
Frank~J. {Qu}, Blake~D. {Sherwin}, Mathew~S. {Madhavacheril}, Dongwon {Han},
  Kevin~T. {Crowley}, Irene {Abril-Cabezas}, Peter A.~R. {Ade}, Simone {Aiola},
  et~al.
\newblock {The Atacama Cosmology Telescope: A Measurement of the DR6 CMB
  Lensing Power Spectrum and its Implications for Structure Growth}.
\newblock {\em arXiv e-prints}, page arXiv:2304.05202, April 2023.

\bibitem{microwave:Wu2019}
W.~L.~K. {Wu}, L.~M. {Mocanu}, P.~A.~R. {Ade}, A.~J. {Anderson}, J.~E.
  {Austermann}, J.~S. {Avva}, J.~A. {Beall}, A.~N. {Bender}, et~al.
\newblock {A Measurement of the Cosmic Microwave Background Lensing Potential
  and Power Spectrum from 500 deg$^{2}$ of SPTpol Temperature and Polarization
  Data}.
\newblock {\em Astrophys. J.}, 884(1):70, October 2019.

\bibitem{microwave:modes}
Douglas {Scott}, Dagoberto {Contreras}, Ali {Narimani}, and Yin-Zhe {Ma}.
\newblock {The information content of cosmic microwave background
  anisotropies}.
\newblock {\em J. Cosmology Astropart. Phys.}, 2016(6):046, Jun 2016.

\bibitem{microwave:KnoxSong}
Lloyd {Knox} and Yong-Seon {Song}.
\newblock {Limit on the Detectability of the Energy Scale of Inflation}.
\newblock {\em Phys. Rev. Lett.}, 89(1):011303, Jul 2002.

\bibitem{microwave:Kesdenetal}
Michael {Kesden}, Asantha {Cooray}, and Marc {Kamionkowski}.
\newblock {Separation of Gravitational-Wave and Cosmic-Shear Contributions to
  Cosmic Microwave Background Polarization}.
\newblock {\em Phys. Rev. Lett.}, 89:011304, Jul 2002.

\bibitem{microwave:HirataSeljak}
Christopher~M. {Hirata} and Uro{\v{s}} {Seljak}.
\newblock {Reconstruction of lensing from the cosmic microwave background
  polarization}.
\newblock {\em Phys. Rev.}, D68(8):083002, Oct 2003.

\bibitem{microwave:PlanckComponents}
{Planck Collab.\ 2013 Results XII}.
\newblock {Planck 2013 results. XII. Diffuse component separation}.
\newblock {\em Astron. Astrophys.}, 571:A12, November 2014.

\bibitem{microwave:Planck2015ComponentsII}
{Planck Collab.\ 2015 Results X}.
\newblock {Planck 2015 results. X. Diffuse component separation: Foreground
  maps}.
\newblock {\em Astron. Astrophys.}, 594:A10, September 2016.

\bibitem{microwave:Planck2018Components}
{Planck Collab.\ 2018 Results IV}.
\newblock {Planck 2018 results. IV. Diffuse component separation}.
\newblock {\em Astron. Astrophys.}, 641:A4, September 2020.

\bibitem{microwave:gold11}
B.~{Gold}, N.~{Odegard}, J.~L. {Weiland}, R.~S. {Hill}, A.~{Kogut}, C.~L.
  {Bennett}, G.~{Hinshaw}, X.~{Chen}, et~al.
\newblock {Seven-year Wilkinson Microwave Anisotropy Probe (WMAP) Observations:
  Galactic Foreground Emission}.
\newblock {\em Astrophys. J. Supp.}, 192:15, February 2011.

\bibitem{microwave:PlanckDust}
{Planck Collab.\ Interm.\ Results XXX}.
\newblock {Planck intermediate results. XXX. The angular power spectrum of
  polarized dust emission at intermediate and high Galactic latitudes}.
\newblock {\em Astron. Astrophys.}, 586:A133, February 2016.

\bibitem{microwave:millea11}
M.~{Millea}, O.~{Dor{\'e}}, J.~{Dudley}, G.~{Holder}, L.~{Knox}, L.~{Shaw},
  Y.-S. {Song}, and O.~{Zahn}.
\newblock {Modeling Extragalactic Foregrounds and Secondaries for Unbiased
  Estimation of Cosmological Parameters from Primary Cosmic Microwave
  Background Anisotropy}.
\newblock {\em Astrophys. J.}, 746:4, February 2012.

\bibitem{microwave:sunzel80}
{Sunyaev, R.~A.} and {Zeldovich, \relax{Ya}.~B.}
\newblock {Microwave background radiation as a probe of the contemporary
  structure and history of the universe}.
\newblock {\em Ann. Rev. Astron. Astrophys.}, 18:537--560, 1980.

\bibitem{microwave:SPTSZ}
R.~{Williamson}, B.~A. {Benson}, F.~W. {High}, K.~{Vanderlinde}, P.~A.~R.
  {Ade}, K.~A. {Aird}, K.~{Andersson}, R.~{Armstrong}, et~al.
\newblock {A Sunyaev-Zel'dovich-selected Sample of the Most Massive Galaxy
  Clusters in the 2500 deg$^{2}$ South Pole Telescope Survey}.
\newblock {\em Astrophys. J.}, 738:139, September 2011.

\bibitem{microwave:ACTSZ2}
M.~{Hilton}, C.~{Sif{\'o}n}, S.~{Naess}, M.~{Madhavacheril}, M.~{Oguri},
  E.~{Rozo}, E.~{Rykoff}, T.~M.~C. {Abbott}, et~al.
\newblock {The Atacama Cosmology Telescope: A Catalog of >4000
  Sunyaev{\textendash}Zel{\textquoteright}dovich Galaxy Clusters}.
\newblock {\em Astrophys. J. Supp.}, 253(1):3, March 2021.

\bibitem{microwave:PlanckSZ}
{Planck Collab.\ Early Results VIII}.
\newblock {Planck early results. VIII. The all-sky early Sunyaev-Zeldovich
  cluster sample}.
\newblock {\em Astron. Astrophys.}, 536:A8, December 2011.

\bibitem{microwave:PlanckSZCosmology}
{Planck Collab.\ 2013 Results XX}.
\newblock {Planck 2013 results. XX. Cosmology from Sunyaev-Zeldovich cluster
  counts}.
\newblock {\em Astron. Astrophys.}, 571:A20, November 2014.

\bibitem{microwave:DeHaan16}
T.~{de Haan}, B.~A. {Benson}, L.~E. {Bleem}, S.~W. {Allen}, D.~E. {Applegate},
  M.~L.~N. {Ashby}, M.~{Bautz}, M.~{Bayliss}, et~al.
\newblock {Cosmological Constraints from Galaxy Clusters in the 2500
  Square-degree SPT-SZ Survey}.
\newblock {\em Astrophys. J.}, 832:95, November 2016.

\bibitem{microwave:Planck2018NG}
{Planck Collab.\ 2018 Results IX}.
\newblock {Planck 2018 results. IX. Constraints on primordial non-Gaussianity}.
\newblock {\em Astron. Astrophys.}, 641:A9, September 2020.

\bibitem{microwave:Planck2015IandS}
{Planck Collab.\ 2015 Results XVI}.
\newblock {Planck 2015 results. XVI. Isotropy and statistics of the CMB}.
\newblock {\em Astron. Astrophys.}, 594:A16, September 2016.

\bibitem{microwave:PlanckInfl}
{Planck Collab.\ 2013 Results XXII}.
\newblock {Planck 2013 results. XXII. Constraints on inflation}.
\newblock {\em Astron. Astrophys.}, 571:A22, November 2014.

\bibitem{microwave:Planck2015Infl}
{Planck Collab.\ 2015 Results XX}.
\newblock {Planck 2015 results. XX. Constraints on inflation}.
\newblock {\em Astron. Astrophys.}, 594:A20, September 2016.

\bibitem{microwave:PlanckDEMG}
{Planck Collab.\ 2015 Results XIV}.
\newblock {Planck 2015 results. XIV. Dark energy and modified gravity}.
\newblock {\em Astron. Astrophys.}, 594:A14, September 2016.

\bibitem{microwave:Reion16}
{Planck Collab.\ Interm.\ Results XLVI}.
\newblock {Planck intermediate results. XLVI. Reduction of large-scale
  systematic effects in HFI polarization maps and estimation of the
  reionization optical depth}.
\newblock {\em Astron. Astrophys.}, 596:A107, December 2016.

\bibitem{microwave:fan06}
X.~{Fan}, C.~L. {Carilli}, and B.~{Keating}.
\newblock {Observational Constraints on Cosmic Reionization}.
\newblock {\em Ann. Rev. Astron. Astrophys.}, 44:415--462, September 2006.

\bibitem{microwave:Mason18}
C.~A. {Mason}, T.~{Treu}, M.~{Dijkstra}, A.~{Mesinger}, M.~{Trenti},
  L.~{Pentericci}, S.~{de Barros}, and E.~{Vanzella}.
\newblock {The Universe Is Reionizing at z {\ensuremath{\sim}} 7: Bayesian
  Inference of the IGM Neutral Fraction Using Ly{\ensuremath{\alpha}} Emission
  from Galaxies}.
\newblock {\em Astrophys. J.}, 856(1):2, March 2018.

\bibitem{microwave:PlanckParity}
{Planck Collab.\ Interm.\ Results XLIX}.
\newblock {Planck intermediate results. XLIX. Parity-violation constraints from
  polarization data}.
\newblock {\em Astron. Astrophys.}, 596:A110, December 2016.

\bibitem{microwave:Komatsu2022}
Eiichiro {Komatsu}.
\newblock {New physics from the polarized light of the cosmic microwave
  background}.
\newblock {\em Nature Reviews Physics}, 4(7):452--469, May 2022.

\bibitem{microwave:KamKos}
M.~{Kamionkowski} and A.~{Kosowsky}.
\newblock {The Cosmic Microwave Background and Particle Physics}.
\newblock {\em Ann. Rev. Nucl. Part. Sci.}, 49:77--123, 1999.

\bibitem{microwave:PlanckMagnetic}
{Planck Collab.\ 2015 Results XIX}.
\newblock {Planck 2015 results. XIX. Constraints on primordial magnetic
  fields}.
\newblock {\em Astron. Astrophys.}, 594:A19, September 2016.

\bibitem{microwave:PlanckConstants}
{Planck Collab.\ Interm.\ Results XXIV}.
\newblock {Planck intermediate results. XXIV. Constraints on variations in
  fundamental constants}.
\newblock {\em Astron. Astrophys.}, 580:A22, August 2015.

\end{thebibliography}

\end{document}